\documentclass{crtbt_modif} 

\Date{Dec 23, 2003} 
\PSLogo{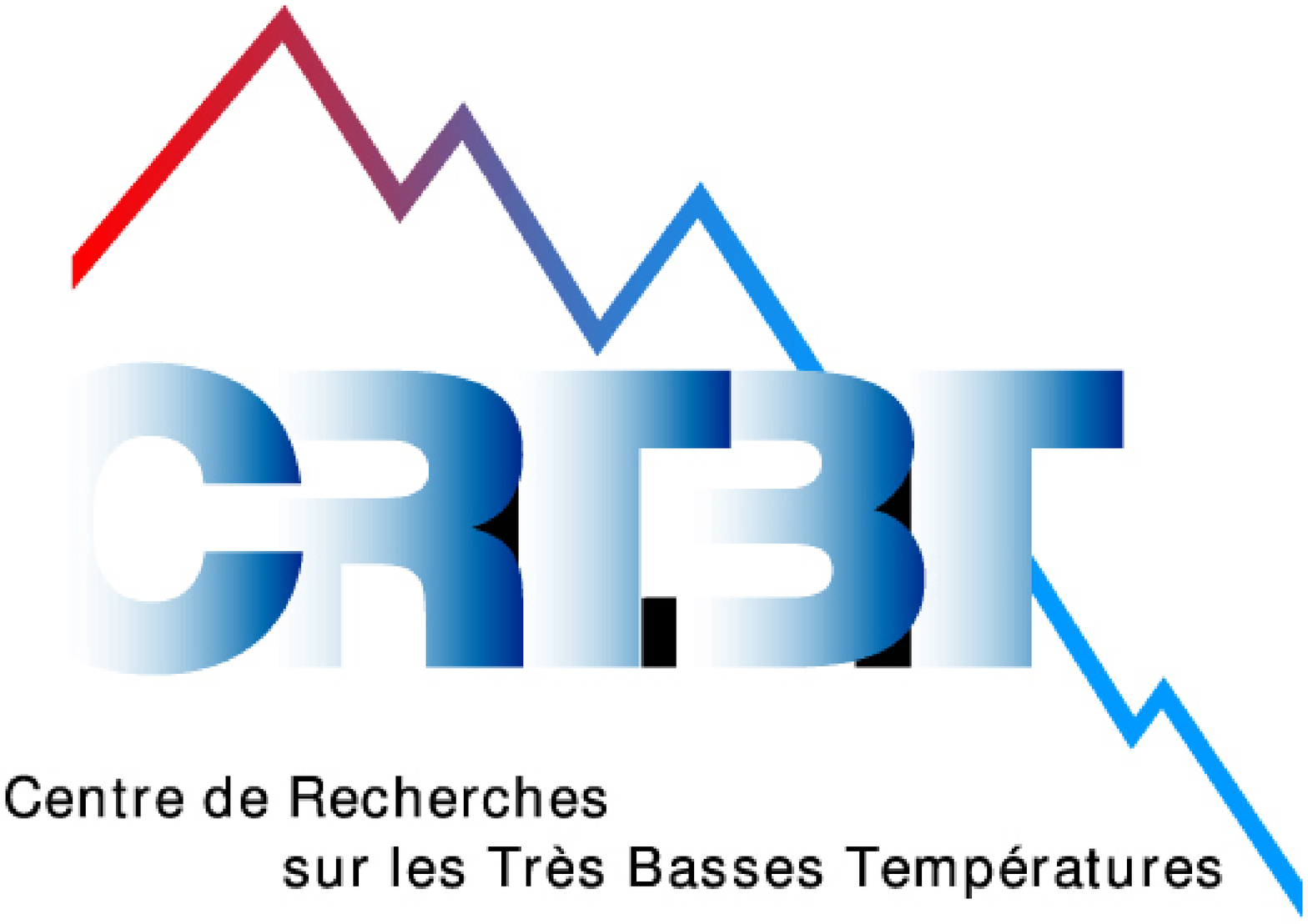}
\LogoHeight{1.5cm}
\Author{P.-E. Roche$^{1,2}$, B. Derrida$^3$ and B. Dou\c{c}ot$^4$} 
\Title{Mesoscopic Full Counting Statistics\\ and Exclusion models\\ } 

\AdresseBis{$^2$ Laboratoire de Physique de la Mati\`ere Condens\'ee\\ Ecole Normale Sup\'erieure, 24 rue Lhomond, 75231 Paris Cedex 05, France\\ \\
$^3$ Laboratoire de Physique Statistique\\ Ecole Normale Sup\'erieure, 24 rue Lhomond, 75231 Paris Cedex 05, France\\ \\
$^4$ Laboratoire de Physique Th\'eorique et des Hautes Energies\\Universit\'e Denis Diderot, 4 place Jussieu, 75252 Paris Cedex 05, France} 

\begin{document} 
\Maketitle 

\Summary{
We calculate the distribution of current fluctuations in two simple exclusion models. Although these models are classical, we recover even for small systems such as a simple or a double barrier, the same distibution of current as given by traditionnal formalisms for quantum mesoscopic conductors. Due to their simplicity, the full counting statistics in exclusion models can be reduced to the calculation of the largest eigenvalue of a matrix, the size of which is the number of internal configurations of the system.
As examples, we derive the shot noise power and higher order statistics of current fluctuations (skewness, full counting statistics, ....) of various conductors, including multiple barriers, diffusive islands between tunnel barriers and diffusive media. A special attention is dedicated to the third cumulant, which experimental measurability has been demonstrated lately. }


\section{Introduction\label{intro}}

A constant voltage difference across a conductor drives an electrical current which will always fluctuate around its mean value. Fluctuations result from random microscopic processes (thermal relaxation, scattering, tunneling...) undergone by the charge carriers. These fluctuations can be considered as an undesirable noise but also as a rich signature of the basic transport mechanisms occurring in the conductor. This second perspective has concentrated much attention in the mesoscopic community over the last decade\cite{blanter:2000}.

In our previous paper\cite{derrida2004} we gave evidence that the statistics of current fluctuations in 
a large classical model, the symmetric exclusion process, 
are identical to the ones derived for quantum mesoscopic conductors\cite{levitov:1996}. Here,
we show that exclusion models allow also to recover the current fluctuations
of small systems such as a single or a double barrier, even in the ballistic limit. 

In the present paper, we develop a classical approach 
to derive the statistics of current fluctuations in mesoscopic conductors (``quantum conductors'') and more generally in conductors smaller than the electronic inelastic mean free path and for some inelastic conductors. Solving the current statistics problem is reduced to finding the largest eigenvalue of a modified evolution matrix, later called the \textit{counting matrix}. We recover and generalize
the well known current statistics for a few mesoscopic systems. Our description is based on the \textit{exclusion process models}, which have been widely studied in statistical physics 
and probability theory\cite{spohn:1991,derrida:2001asep,liggett:1999}.
The main benefits of this approach are its conceptual and  analytical simplicity.

In the remaining part of this introduction section, we briefly 
recall the traditional approaches for mesoscopic transport (section~\ref{formalism}) and the basic mathematical tools necessary to describe current fluctuations (\ref{math}). Section~\ref{exclumod} presents two exclusion models fitted for condensed matter conductors and the procedure to derive the complete statistics of current fluctuations (later called ``Full Counting Statistics'' or FCS.). In section~\ref{Application}, our exclusion models are used to derive the current statistics of various elementary conductors. 

\subsection{Traditional formalisms for transport in condensed matter physics\label{formalism}}

A number of approaches have already been used to describe 
the FCS in mesoscopic conductors. The Scattering Matrix theory\cite{buttiker:1990,shimizu:1991,buttiker:1992,levitov:1993, muzykantskii, levitov:1996} is well adapted to the
modeling of quantum-mechanically coherent conductors in a regime where electron interaction
effects are sufficiently weak to be neglected. With this strong assumption, this allows
to treat an arbitrary large number of transverse conduction channels, which yield
independent contributions to the current statistics. This approch, combined with results
from random matrix theory for scattering matrices\cite{beenakker:1997} 
has lead to precise predictions
for the FCS of a disordered conductor in the diffusive regime\cite{lee}. A more direct microscopic
treatment of disordered systems relies on the Keldysh technique\cite{keldysh} to construct the non-equilibrium
density matrix of the steady state at finite current. Disorder averaging is then performed
using a non-linear Sigma model representation\cite{gutman:2002cumulants}. A rather general
\textit{circuit theory} has been constructed to account for the influence of an arbitrary environment,
described in terms of an equivalent circuit, on the measured fluctuations of a mesoscopic conductor 
\cite{nazarov:1995,nazarov:2002}.
Semi-classical descriptions, based on the Boltzmann-Langevin model\cite{shulman,Gantsevich} have also been used to derive the first four cumulants of current fluctuations in a diffusive medium\cite{nagaev:1992,nagaev:2002}. Other semi-classical approaches focused on high order statistics and FCS of a double tunnel barrier\cite{deJong:1996double}, chaotic cavities\cite{pilgram} and diffusive media\cite{roche:2002exclusion, derrida2004}.  The semi-classical results are the same as the ones obtained with the corresponding quantum conductor model.

The exclusion models discussed in this paper represent an extreme  
semi-classical approximation~: the only quantum rule which is preserved is Pauli exclusion principle. In particular, electrons have no phase and don't interfere.

\subsection{Mathematical formalism for Current fluctuations\label{math}}

If $q_t$ is the algebraic charge which flows accross a section during time $t$, the fluctuations of current  $I=q_t / t$ depends in principle on the duration $t$ chosen to measure $I$.
In practice
the long time response of the measuring electronics apparatus sets a lower bound on $t$~: this bound is most often decades larger than all the physical times experienced by charge carriers (diffusion time, dwell time, coherence times in the conductor and in the electrodes,...). Thus, experiments correspond to the $t\rightarrow \infty$ limit, often called the zero-frequency limit in the shot noise literature. In this limit, the choice of the cross-section is irrelevant since the maximum charge accumulation between two different cross-sections is finite, at least in a conductor connected to two electrodes only.

In a conductor smaller than the inelastic mean free path, carriers do not undergo inelastic collisions.
 It is therefore reasonable in many situations to neglect interaction effects on such small
length scales. Equivalently, we may then assume that these charge carriers
remain on independent energy levels\cite{blanter:2000}. 
Consequently the statistics of the total current will consist in a summation of independent random variables corresponding to different energy levels. In the following, to keep equations free of elementary-charge prefactors, we focus on carriers counting rather than charge counting. In addition, we will call this generic charge carrier an \textit{electron}.

If $P_{t,\epsilon}(Q)$ is the probability that $Q$ electrons have been transfered at the energy level $\epsilon$ during a time interval $t$, one can fully characterize the counting statistics by the cumulant generating function~:
\begin{equation}
{\cal S}_{t,\epsilon}(z) = \ln \left[ \sum_{Q=-\infty}^\infty  P_{t,\epsilon}(Q)   \ z^{Q} \right] =\ln( \overline{z^Q} )
\label{St}
\end{equation}
or equivalently by cumulants (the $n^{th}$ order one is written here)~:
\begin{equation}
C_n(t,\epsilon)=\frac{\partial^{n}{\cal S}_{t,\epsilon}(z)}{\partial(\ln z)^{n}}
\label{Ci}
\end{equation}
we have in particular~: $C_1=\bar{Q}$ , ~ $C_2=\overline{(Q-\bar{Q})^2}$,~ $C_3=\overline{(Q-\bar{Q})^3}$ , $C_4(\epsilon)=\overline{(Q-\bar{Q})^4}-3\overline{(Q-\bar{Q})^2}^2$,~...

For a given conductor, the current at an energy level $\epsilon$ only depends on the boundary conditions, that is the fillings $\rho_{_L}(\epsilon)$ and $\rho_{_R}(\epsilon)$ of the left and right electrodes (or ``reservoirs'') at both ends of the conductor. If we rewrite the cumulant explicitly as $C_n(t, \epsilon,\rho_{_L},\rho_{_R})$, the cumulants $K_n(t)$ for the whole conductor are given by
\begin{equation}
\label{eq:mu}
K_n(t)=\int{C_n(t, \epsilon,\rho_{_L}(\epsilon),\rho_{_R}(\epsilon))\;n(\epsilon)\;d\epsilon}
\label{integration}
\end{equation}
where $n(\epsilon)$ is the density of energy levels in the conductor. Likewise, the cumulant generating function for the whole conductor can be derived with the same type of summation. For comparison with experiments Fermi-Dirac distributions are imposed in the left and right electrodes~:
\begin{equation}
\label{eq:FD}
\rho_{_L}(\epsilon)=\frac{1}{1+e^{\frac{\epsilon-eV}{k_{B}T}}} ~~~;~~~ \rho_{_R}(\epsilon)=\frac{1}{1+e^{\frac{\epsilon}{k_{B}T}}}
\label{FermiDirac}
\end{equation}
With such fillings, the $K_n(t)$ are function of the driving voltage normalized by temperature $eV/k_{B}T$. 
The $k_{B}T\gg eV$ limit corresponds to the Johnson-Nyquist thermal noise and the opposite limit to pure shot noise. In this paper, the integration Eq.~(\ref{integration}) over $\epsilon$ will be estimated assuming that $n(\epsilon)$ and $C_n(t,\epsilon,\rho_{_L},\rho_{_R})$ are independent of $\epsilon$. This assumption is quite reasonable,
since in most cases the Fermi energy in the reservoirs is much larger than both the thermal
energy window $k_{B}T$ and the driving energy $eV$.

The electrical conductance $G$ and the current noise power density $S_{I}$ are proportional to $K_1$ and $K_2$~:
\begin{equation}
G=\overline{I}/V=eK_1/Vt ~~~;~~~S_{I}=2\int_{}^{}\overline{\delta I(\tau)\delta I(0)} d\tau=2e^2K_2/t
\label{defGSI}
\end{equation}
where $\delta I(\tau)=I(\tau)-\overline{I}$ is the current fluctuation at time $\tau$. 
The time scale of the model dynamics can be chosen arbitrarily since this only changes the prefactor of the cumulant generating function. 
The transport mechanism is characterized by the cumulants $C_2$, $C_3$,... (or $K_2$, $K_3$,...) normalized by $C_1$ (or $K_1$). In particular, we will focus on the normalized shot noise power 
\begin{equation}
F=S_I/2e\overline{I}=K_2/K_1
\label{FanoFactorGene}
\end{equation}
and the normalized skewness 
\begin{equation}
F_3=K_3/K_1
\end{equation}

\section{Exclusion Models\label{exclumod}}

In this section, we first present an exclusion model mostly adapted to the modeling of nearly ballistic conductors. 
We call it the \textit{counter-flows exclusion model} because the two directions of propagation of electrons found in a 1D conductor are explicitly considered. Conductors with a low transmission efficiency, such as tunnel barriers or diffusive media, can sometimes be described by a simpler exclusion model, presented in the \textit{tunnel exclusion model} section. Many systems studied in the exclusion, hopping-model  and sequential-tunneling literatures are directly relevant to this latter category of conductors. These models describe independent particles, apart for
the exclusion constraint which represents the effect of the Pauli principle.

\subsection{The Counter-Flows Exclusion model}

The counter-flows model is inspired from the Landauer\cite{lesovik:1989,Yurke,buttiker:1990,landauer:1991} picture of conductors~: at zero temperature, electrons are injected periodically from the reservoirs to the conductor. 
This assumption seems a good enough modeling to account for the FCS in the $t\rightarrow \infty$ limit. 
Indeed, the predictions of the model would remain unchanged if the variance 
$\overline{(N_t-\overline{N_t})^{2}}$ of the number of injection attempts $N_t$ 
during a time interval $t$, is only sublinear in $t$. This later property follows 
from the Pauli exclusion in degenerate electrodes which imposes an anti-correlation between injection events\cite{levitov:1993, levitov:1996}. The experimental validation of Landauer approach\cite{reznikov:1995,Kumar:1996,blanter:2000}
justifies a posteriori this nearly-periodic injection model. While in the sample,
these charge carriers may undergo internal scattering on localized barriers and 
finally are either reflected or transmitted to electrodes at both ends of the conductors. 
The Pauli exclusion principle is fulfilled at each stage during the system evolution. 
In the original model, electrons are described by wave-packets and phase coherence is preserved during the scattering but alternative versions of this model dropped the phase information (for example, see\cite{deJong:1995,liu}).\\

\begin{figure}
\centerline{\includegraphics[width=5cm]{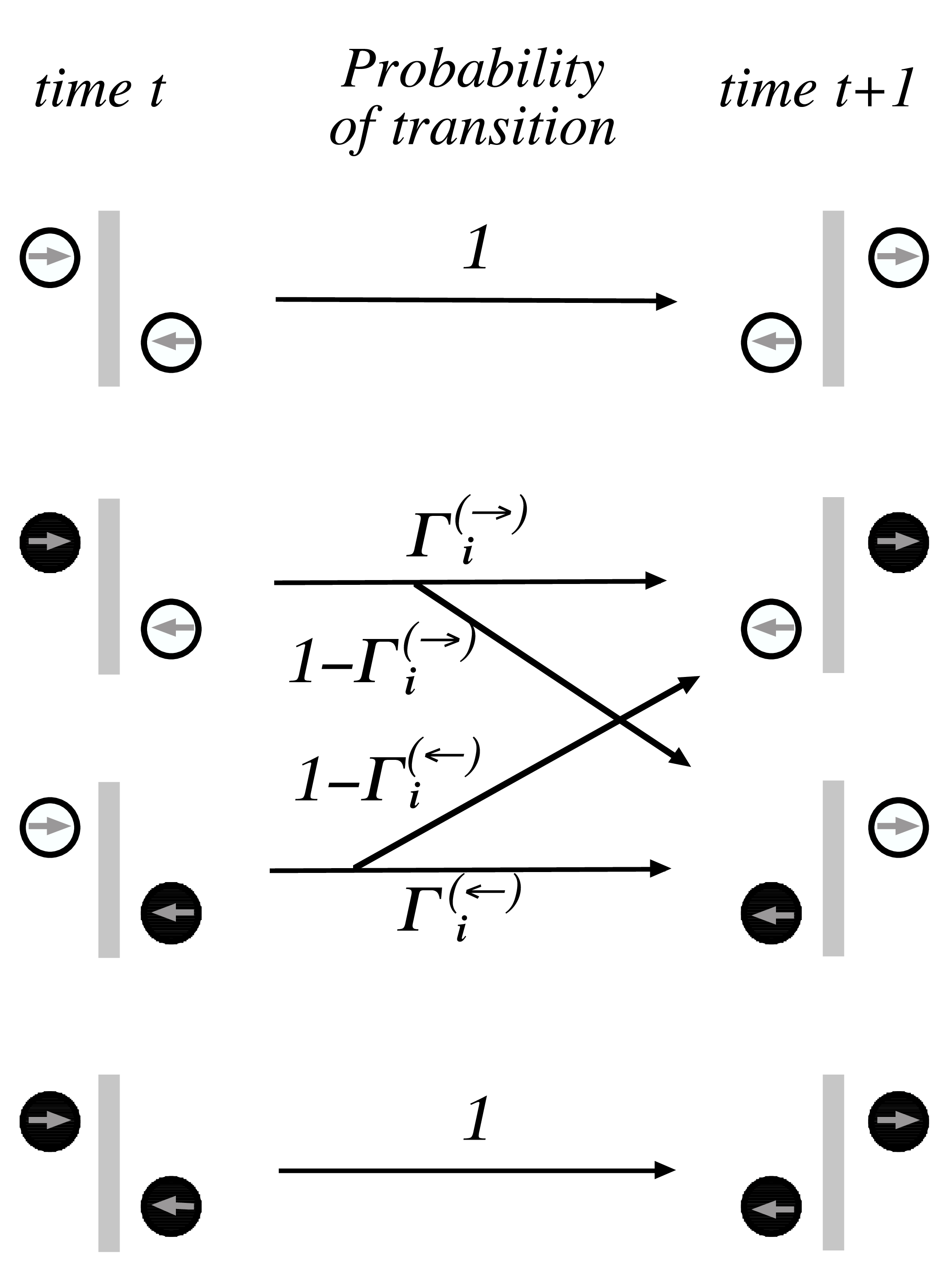}}
\caption{Counter-flows model. Probability of evolution for the various configurations of electrons reaching the $i^{th}$ barrier at time $t$.
}
\label{fig:counterFlow2}
\end{figure}

More precisely, the $2N$ sites counter-flows model consists in $N+1$ barriers, each characterized by 2 transmission probabilities $\Gamma_{i}^{(\rightarrow)}$ (from left to right) and $\Gamma_{i}^{(\leftarrow)}$ (from right to left) where $i$ is the index of the barrier increasing from left to right ($1\leq i\leq N+1$). Between two consecutive barriers, 2 sites are available for at most 2 electrons propagating in opposite directions. So a configuration at time $t$ is characterized by $2N$ binary variables
$\tau_{i}^{(\rightarrow)}(t)$ and $\tau_{i}^{(\leftarrow)}(t)$ for $1\leq i\leq N$;
$\tau_{i}^{(\rightarrow)}(t)$ (respectively $\tau_{i}^{(\leftarrow)}(t)$) is equal to 1 if
an electron propagating to the right (respectively to the left) is present at site $i$ at time $t$.
Time is discrete and at each time step, electrons are transmitted through one barrier to the next site, unless a back-scattering occurs on the barrier. By definition of the dynamics of the model,
$\tau_{i+1}^{(\rightarrow)}(t+1)$ and  $\tau_{i}^{(\leftarrow)}(t+1)$ depend only on
$\tau_{i}^{(\rightarrow)}(t)$ and on $\tau_{i+1}^{(\leftarrow)}(t)$, and the (classical)
transition probabilities are given in figure \ref{fig:counterFlow2}. This allows for
a simultaneous update of all occupancies, even in the presence of backscattering on barriers.

\begin{figure}
\centerline{\includegraphics[width=8cm]{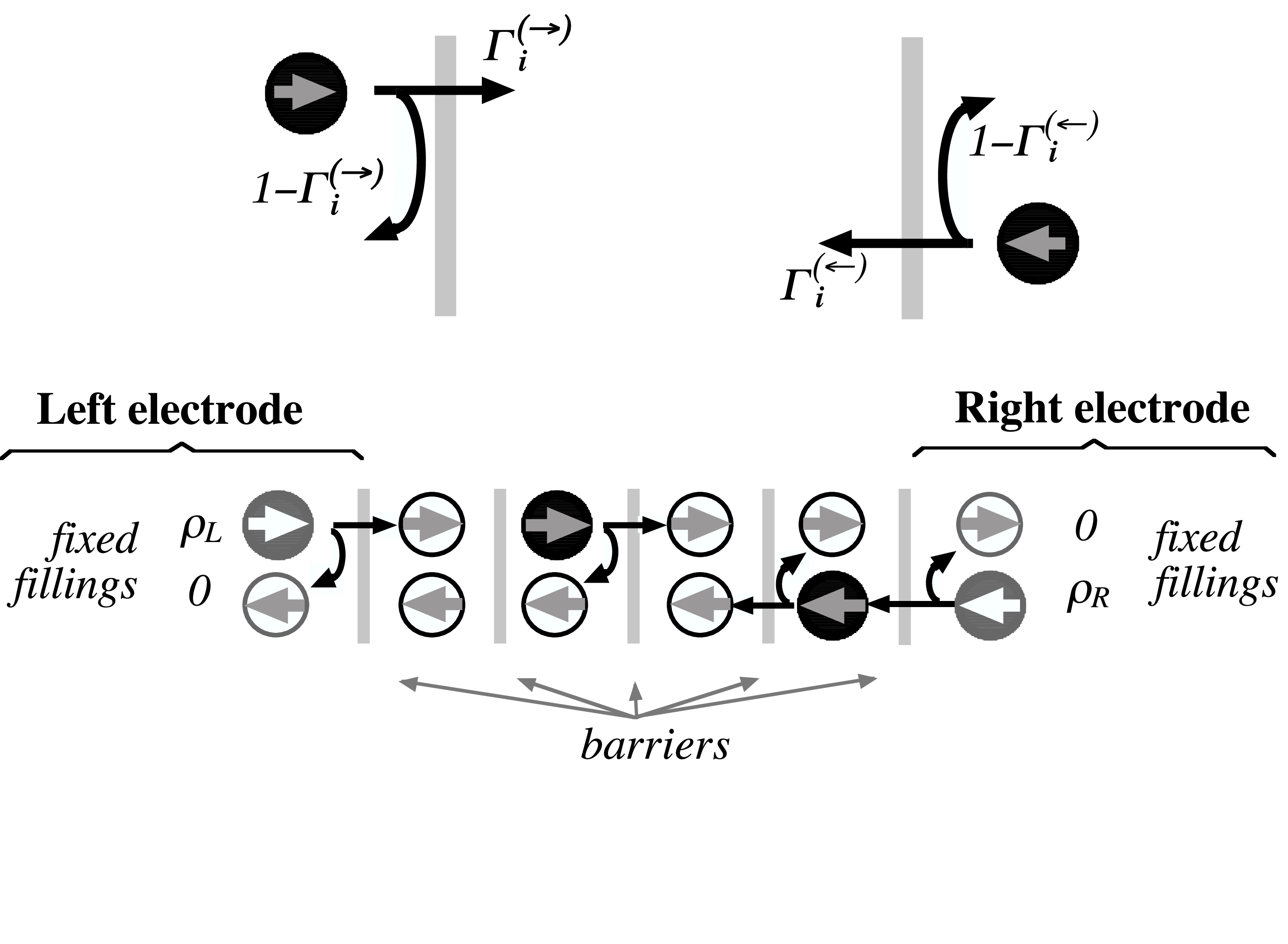}}
\caption{Counter-flows model. Upper Fig.~: Transmission and reflection probabilities for the $i^{th}$ barrier, assuming no conflict with the exclusion principle. Lower Fig.~: The counter-flows model for $N=4$. The white circles represent empty sites, the black disks are electrons, the gray disks stand for sites with a fixed filling probability  and the arrows indicate the direction of propagation associated with each site.
}
\label{fig:counterFlow}
\end{figure}

At the boundaries of the conductors, each electrode is modeled by 2 sites, the occupation states of which are re-set before each time step. The site corresponding to an electron propagating into the conductor is re-filled with probability $\rho_{_L}$ (left electrode) or $\rho_{_R}$ (right electrode) and the site accessible to the electron leaving the conductor is re-emptied at each time step (see Fig.\ref{fig:counterFlow}). 
The densities $\rho_{_L}$ and $\rho_{_R}$ are given by Fermi-Dirac distributions (Eq.~(\ref{eq:FD})). After this reset, the one-time-step evolution follows the same transmission/back-scattering rule that holds in the bulk of the conductor.

On modeling real conductors, the barriers can represent junctions (between two different materials for example),  scattering centers (impurities, structural defects, ...) or even inelastic processes (phonon or photon-assisted hopping, emission of phonon or photon, ...). The model parameters $N$, $\Gamma_{i}^{(\leftarrow)}$ and 
$\Gamma_{i}^{(\rightarrow)}$ are related to the corresponding physical quantities such as tunneling probabilities or  scattering cross-sections.

\subsection{The Tunnel Exclusion Model\label{tunnelExclusion}}

\begin{figure}
\centerline{\includegraphics[width=8cm]{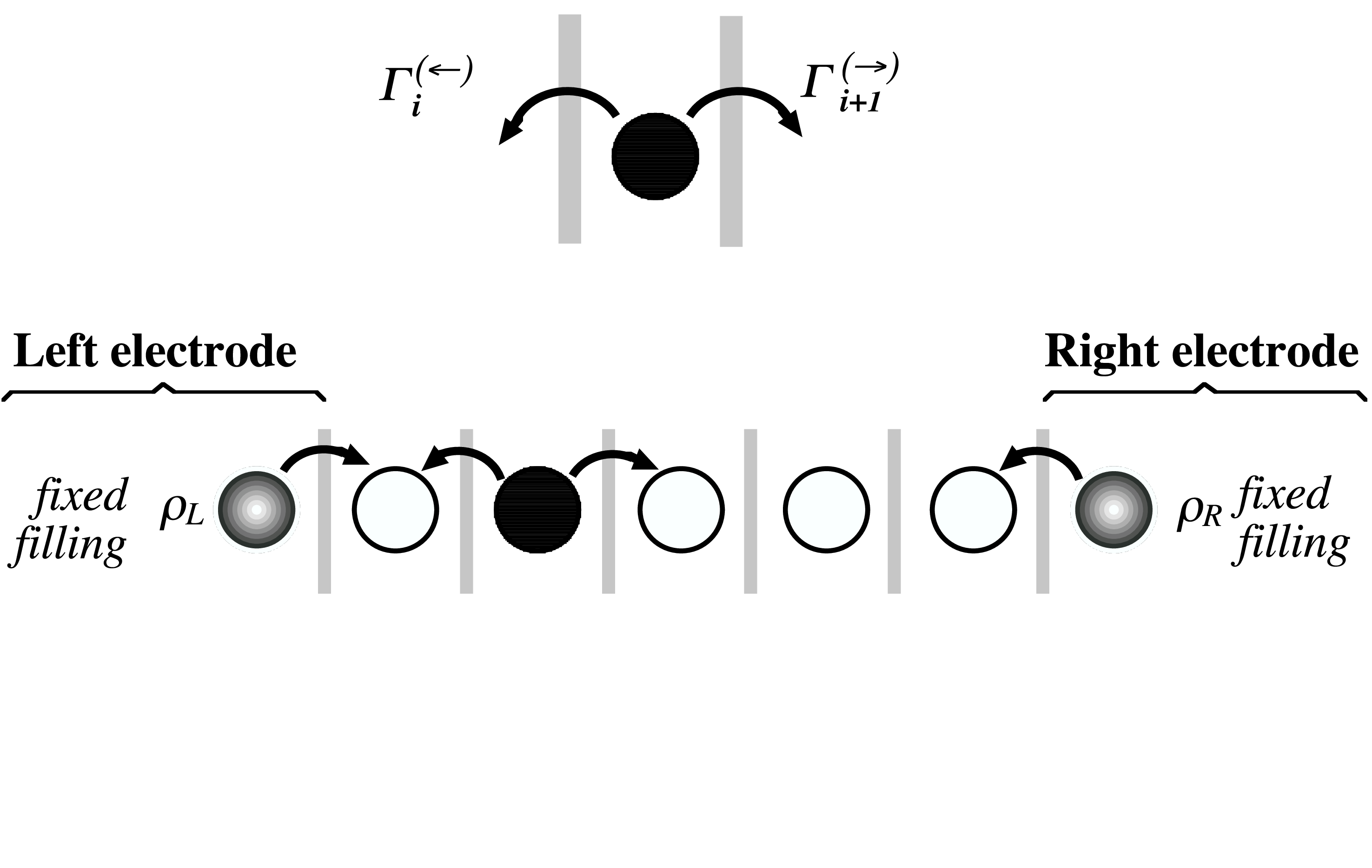}}
\caption{Tunnel model. Upper Fig.~: Tunneling probabilities across the $i^{th}$ and $(i+1)^{th}$ barriers for an electron located between them, assuming no conflict with the exclusion principle. Lower Fig.~: Tunnel exclusion model for N=5 sites.
}
\label{fig:tunnelFlow}
\end{figure}

It is useful to note that the counter-flows exclusion model may be decomposed into two
independent stochastic models. Let us define new variables $\sigma_{i}(t)$ and
$\sigma_{i}'(t)$ such that $\sigma_{i}(t)=\tau_{i}^{(\rightarrow)}(t)$ if
$i$ and $t$ have the same parity and else $\sigma_{i}(t)=\tau_{i}^{(\leftarrow)}(t)$.
In a similar way, $\sigma_{i}'(t)=\tau_{i}^{(\leftarrow)}(t)$
if $i$ and $t$ have the same parity and else
$\sigma_{i}'(t)=\tau_{i}^{(\rightarrow)}(t)$.
From the definition on the model, the random variables $\sigma_{i}$ are completely
decoupled from the $\sigma_{i}'$ variables.
It turns out that in the limit of small transmission probabilities, the dynamics of each  
of these two ensembles of binary variables may be formulated in terms of a simpler lattice model,
that we shall call the \textit{tunnel exclusion model}. The elementary time-step of the latter
model involves two steps in the former one. This has the advantage that in the limit of
a vanishing transmission probability, the configuration of $\sigma_{i}$'s does not
evolve in time.  For each of these two independent submodels, expanding the evolution
of this reduced system to first order in transmission probabilities, and taking the continuous 
time limit, we get the model which definition is sketched on Fig.\ref{fig:tunnelFlow}.
In this case, the quantities
$\Gamma_{i}^{(\rightarrow)}$ and $\Gamma_{i}^{(\leftarrow)}$ become 
the probabilities per time unit of tunneling across the $i^{th}$ barrier from left to right and vice-versa, provided that the target site is empty. 
Each electrode is modeled by a single site, the occupation of which 
is reset to $\rho_{_L}$ (left electrode) or $\rho_{_R}$ (right electrode) before each time step. The fillings $\rho_{_L}$ and  $\rho_{_R}$ are given by Fermi-Dirac distributions (See Eq.~(\ref{eq:FD})).\\


\begin{figure}
\centerline{\includegraphics[width=8cm]{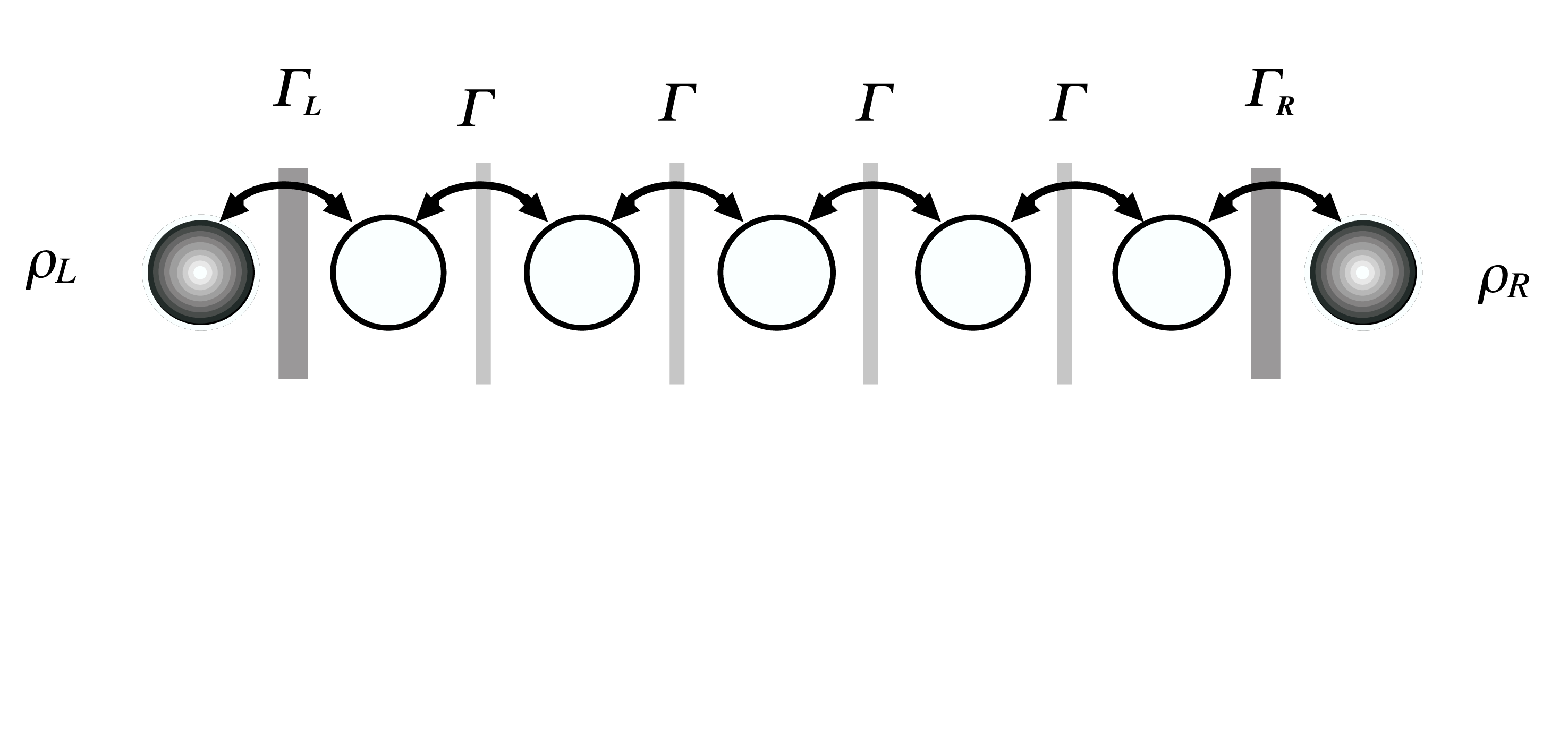}}
\caption{SSEP model for N=5 sites. $\Gamma_{_L}$, $\Gamma$ and $\Gamma_{_R}$ are the tunneling probabilities and $\rho_{_L}$, $\rho_{_R}$ the electrodes' fixed fillings .
}
\label{fig:SSEP}
\end{figure}

A special choice of the tunneling probabilities is the \textit{Symmetric Simple Exclusion Process} or SSEP\cite{spohn} (see~Fig.\ref{fig:SSEP}) for which the the internal barriers are symmetric ($\Gamma_{i}^{(\leftarrow)}=\Gamma_{i}^{(\rightarrow)}$) 
and uniform along the conductor (independent of $i$ for $2\leq i\leq N$). We note this probability $\Gamma$. The two out-most barriers are also modeled with symmetric rates $\Gamma_{_L}=\Gamma_1$ and $\Gamma_{_R}=\Gamma_{N+1}$. Physically they account for the electrical connection between the electrodes and the conductor. 
In the theory of exclusion processes\cite{derrida2004}, one usually represents the reservoirs by injection rates
$\alpha$, $\delta$, and extraction rates $\gamma$, $\beta$ which
give an equivalent description of the boundary conditions if: 
$\alpha=\rho_{_L}\Gamma_{_L}, \delta=\rho_{_R}\Gamma_{_R}, \gamma=(1-\rho_{_L})\Gamma_{_L}, \beta=(1-\rho_{_R})\Gamma_{_R}$.

\subsection{The FCS solving procedure\label{methode}}

\textbf{In the counter-flows model},  the conductor has $2^{2N}$ internal configurations 
${\cal C}=\left\{\tau_{1}^{(\rightarrow)},\tau_{1}^{(\leftarrow)},...,\tau_{N}^{(\rightarrow)},
\tau_{N}^{(\leftarrow)} \right \}$. 
Let $p_t({\cal C})$ be the probability of finding the system in configuration ${\cal C}$ at time $t$.
As the dynamics is a Markov process, the evolution equation for $p_t({\cal C})$ can be written~:
\begin{equation}
p_{t+1}({\cal C})  = \sum_{\cal C'} [M_1({\cal C},{\cal C'}) +   M_0({\cal C},{\cal C'})+M_{-1}({\cal C},{\cal C'})] p_t({\cal C'}) 
\label{evolution_CounterFlow}
\end{equation}
where we have decomposed the evolution matrix into three parts $M_1$, $M_0$ and $M_{-1}$, depending on whether, when  the system jumps from  configuration ${\cal C'}$ to  configuration ${\cal C}$,  the total number of transfered charges increases by $1,0$ or $-1$.

If we define $P_t({\cal C},Q)$ as the probability that the system is in configuration ${\cal C}$ at time $t$ and that $Q$ charges have been transfered, one has~:
\begin{eqnarray}
{P_{t+1}({\cal C},Q) } =  \sum_{\cal C'} M_1({\cal C},{\cal C'}) P_t({\cal C'},Q-1) +   M_0({\cal C},{\cal C'}) P_t({\cal C'},Q)
+ M_{-1}({\cal C},{\cal C'})  P_t({\cal C'},Q+1)
\label{evo2}
\end{eqnarray}
Then the generating functions ${\cal P}_t({\cal C}, z)$ defined  by~:
\begin{equation}
{ \cal P}_t({\cal C}, z) = \sum_{Q=-\infty}^\infty  P_t({\cal C},Q)   \ z^Q
\label{PCzdef}
\end{equation}
 satisfies
\begin{equation}
{{\cal P}_{t+1}({\cal C},z) } =  \sum_{\cal C'}  \left[ z  \  M_1({\cal C},{\cal C'})  +   M_0({\cal C},{\cal C'}) + {1 \over z} M_{-1}({\cal C},{\cal C'}) \right]
  {\cal P}_t({\cal C'},z)
\label{PCz}
\end{equation}
If we introduce $M_z$ that we will call the \textit{counting matrix}, defined by:
\begin{equation}
M_z({\cal C},{\cal C'}) = z  \  M_1({\cal C},{\cal C'})  +   M_0({\cal C},{\cal C'}) + {1 \over z} M_{-1}({\cal C},{\cal C'}) 
\label{counting_matrix_CounterFlow}
\end{equation}
it is clear from Eq.~(\ref{PCz}) that in the long time limit, the cumulant generating function for the total number of transfered charges is~:
\begin{equation}
 {\cal S}_{t}(z)  =\ln( \overline{z^Q} )  = \ln \left[  \sum_{\cal C}{ \cal P}_t({\cal C}, z) \right]  \sim  \ln \left( \nu(z)^ t   \right)  \sim t \ln  \left( \nu(z) \right) 
\label{cgf_CounterFlow}
\end{equation}
where $\nu(z)$ is the largest eigenvalue of the counting matrix $M_z$~~\cite{DerridaLebo:1998,derrida2004}. Due to the fact (see beginning of section \ref{tunnelExclusion}) that the counter-flows model can be decomposed into two decoupled sets of variables, the eigenvalue $\nu(z)$  can in fact be obtained by diagonalizing a $2^Nx2^N$ matrix.

\textbf{In the tunnel model},  the conductor has $2^{N}$ internal configurations and -as previously- we call $p_t({\cal C})$ the probability of finding the system in configuration ${\cal C}$ at time $t$. The time being continuous in this model, one has~:
\begin{equation}
{d p_t({\cal C}) \over dt} = \sum_{\cal C'} [W_1({\cal C},{\cal C'}) +   W_0({\cal C},{\cal C'})+
 W_{-1}({\cal C},{\cal C'})] p_t({\cal C'}) 
\label{evolution_Tunnel}
\end{equation}
where the evolution matrix has been decomposed into three parts $W_1$, $W_0$ and $W_{-1}$, depending on whether when  the system jumps from  configuration ${\cal C'}$ to  configuration ${\cal C}$,  the total number of transfered charges increases by $1,0$ or $-1$. Eq.~(\ref{evolution_Tunnel}) is a continuous time version of Eq.~(\ref{evo2}), the main difference being the diagonal elements of $W_0$ are now all negative.

Following the same procedure as above, we can define the \textit{counting matrix} $W_z$ by~:
\begin{equation}
W_z({\cal C},{\cal C'}) = z  \  W_1({\cal C},{\cal C'})  +   W_0({\cal C},{\cal C'}) + {1 \over z} W_{-1}({\cal C},{\cal C'}) 
\label{counting_matrix_Tunnel}
\end{equation}
and we find the cumulant generating function for the total transfered charge in the long time limit~:
\begin{equation}
 {\cal S}_{t}(z)  =\ln( \overline{z^Q} )   \sim  \ln \left(  e^{\mu(z) \  t }  \right)  \sim t \  \mu(z)
\label{cgf_Tunnel}
\end{equation}
where $\mu(z)$ is the largest eigenvalue of the counting matrix $W_z$. This latter equation can be seen as the first term in the expansion of the corresponding equation obtained in a discrete time approach.

\textbf{Both for the counter-flows and tunnel models}, the FCS is fully determined by the largest eigenvalue of what we called the \textit{counting matrix}. 

The full knowledge of the eigenvalue is not necessary if only the first $n$ cumulants are wanted. In this case, the equation satisfied by the eigenvalues  
$\left| M_z-\nu(z){\cal I} \right|= 0$ (or $\left| W_z-\mu(z){\cal I} \right| =0$) can be solved by a 
perturbation theory and the n$^{th}$ cumulant is obtained from the coefficient of the n$^{th}$ order of the eigenvalue in powers of $\log(z)$ (see Eq.~(\ref{Ci})).
Once the counting matrix is written down, this procedure can be easily performed by an analytical calculation software.



\section{Application to mesoscopic systems\label{Application}}

In the remaining of this paper we derive the FCS or the first cumulants of basic mesoscopic systems. A special attention is dedicated to the current fluctuations skewness (third cumulant), the physical interest(\cite{levitov:2001,gutman:2003HighTemp}) and measurability\cite{reulet2003} of which have been recently  emphasized. Indeed, at high temperature the skewness can reveal information about transport which are not  blurred by thermal fluctuations\cite{levitov:2001,gutman:2003HighTemp}. Some of the results derived are already known and they validate exclusion  modeling for charge conduction in condensed-matter systems. 
The various new results, often derived in a few lines of linear algebra, illustrate the strength of this modeling.\\

\subsection{Asymmetric barrier and Single channel}

The counter-flows model with $N=0$ site is a single barrier between two electrodes of fillings $\rho_{_L}$ and $\rho_{_R}$. Since the system has no internal state, the counting matrix $M_z$ is a scalar. A positive charge transfer (from left to right) will occur with probability $p_+=\rho_{_L} (1-\rho_{_R}) \Gamma^{(\rightarrow)}$ and a negative transfer with probability
$p_-=(1-\rho_{_L}) \rho_{_R} \Gamma^{(\leftarrow)}$ (see Table 1).

\begin{table}
\begin{center}
\begin{tabular}{|c|c|c|}
\hline
& charge& \\
evolution & counting & probability \\
\hline
\includegraphics[width=1cm]{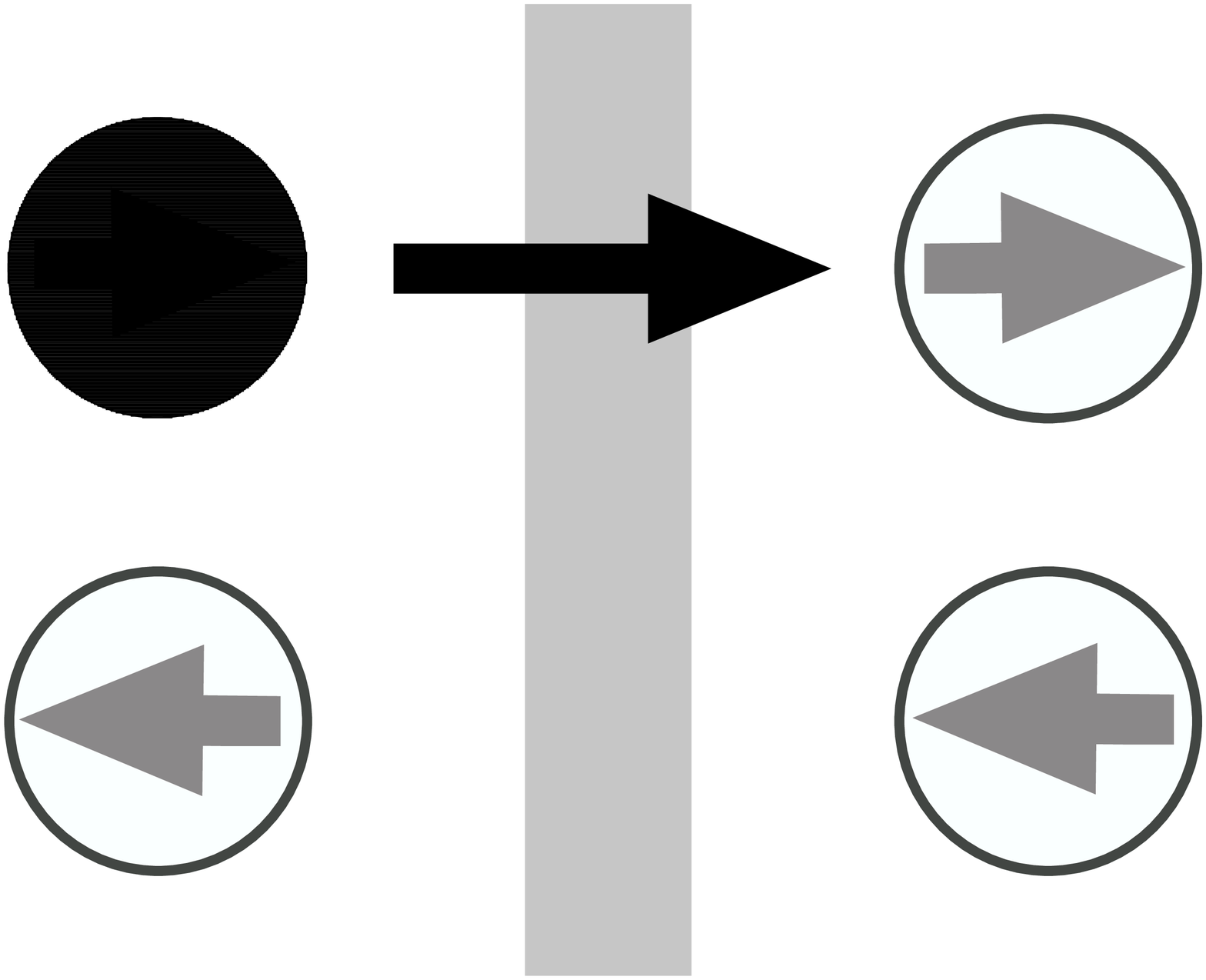}  & increase & $\rho_{_L}(1-\rho_{_R})\Gamma^{(\rightarrow)}$\\ 
\hline
\includegraphics[width=1cm]{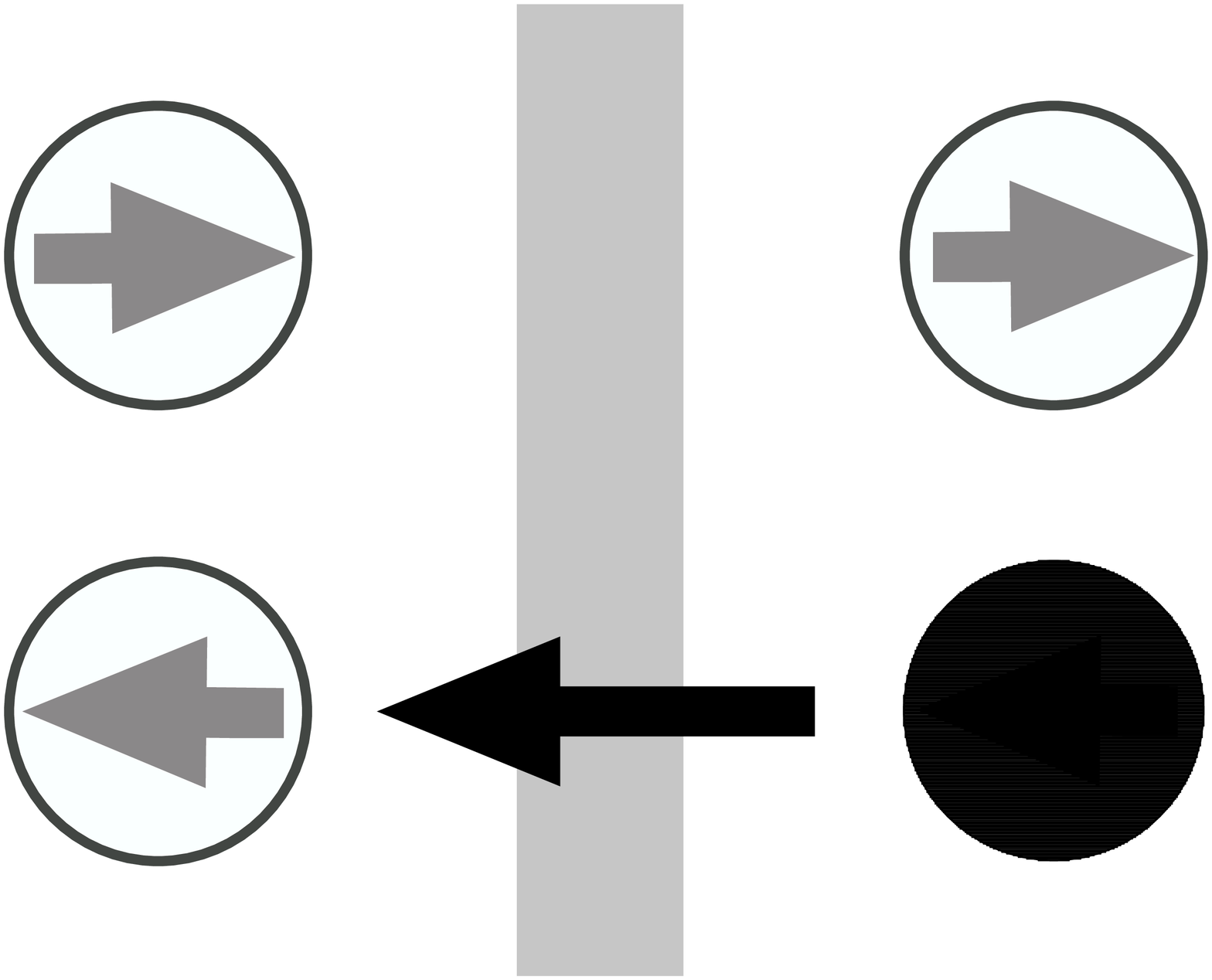}  & decrease & $(1-\rho_{_L})\rho_{_R}\Gamma^{(\leftarrow)}$ \\ 
\hline
others  & unchanged & $1 - \rho_{_L}(1-\rho_{_R})\Gamma^{(\rightarrow)}- (1-\rho_{_L})\rho_{_R}\Gamma^{(\leftarrow)}$ \\ 
\hline
\end{tabular} \\
\caption{Asymmetric Single Barrier for arbitrary fillings $\rho_{_L}$ and $\rho_{_R}$ of the electrodes.}
\end{center}
\end{table}

Following the general procedure for the counter-flows model, we consider the counting ``matrix''~:
\begin{equation}
M_z=p_+\;z \,+\, p_-\;z^{-1} +  ( 1 - p_+ - p_-)
\end{equation}
The logarithm of the largest (and unique) eigenvalue $\nu(z)=M_z$ gives the cumulant generating function $ {\cal S}_{t}(z)$, which fully characterize the FCS of an asymmetrical barrier~:
\begin{equation}
{ {\cal S}_{t}(z) / t}=
 \ln \left\{ \left[ \rho_{_L}\,\left( 1 - \rho_{_R} \right) \, \Gamma^{(\rightarrow)} \right] \, \left( z-1\right) \,+\, \left[ \left( 1 - \rho_{_L} \right) \,\rho_{_R}\, \Gamma^{(\leftarrow)} \right] \, \left( z^{-1} - 1 \right)
\,+ 1 \right\}
\label{singleBaFCS}
\end{equation}

A few particular cases are interesting~:

$\bullet$ The $\Gamma^{(\rightarrow)}, \Gamma^{(\leftarrow)}\rightarrow 1$ limits account for quasi-ballistic barriers. On the opposite case of tunnel barriers ($\Gamma^{(\rightarrow)}, \Gamma^{(\leftarrow)} \ll 1$), the FCS  can be re-estimated from a first order expansion of Eq.~(\ref{singleBaFCS}) or directly with the tunnel exclusion model with $N=0$~:
\begin{equation}
{ {\cal S}_{t}(z) / t}=\mu(z)=W_z=\rho_{_L} (1-\rho_{_R}) \Gamma^{(\rightarrow)} (z-1) + (1-\rho_{_L}) \rho_{_R} 
\Gamma^{(\leftarrow)} (z^{-1}-1)
\end{equation}

$\bullet$ The asymmetric case ($\Gamma^{(\rightarrow)}\neq \Gamma^{(\leftarrow)}$) accounts for inelastic barrier, such as those for which stepping over the barrier requires the emission or assistance of a photon or phonon\cite{levitov:2001}.

For symmetric barriers
\begin{equation}
{\cal T}=\Gamma^{(\rightarrow)}=\Gamma^{(\leftarrow)}
\end{equation}
we recover the important case of a \textit{conduction channel} of  transparency ${\cal T}$ encountered in mesoscopic transport\cite{levitov:1993}. 
This is to be expected from our discussion in section~\ref{quantum}, where we show that for a
single barrier, the counter-flows model can be precisely related to the evolution of the diagonal
part of the density matrix of a quantum many-electron system.
Once the behavior of a single conduction channel has been determined, scattering matrix theory
shows how to reduce 
the problem of interaction-less electronic transport through a quantum constriction into a set of independent symmetric barriers. 
In the zero temperature limit, the only states which contribute to the FCS are those whose energy $\epsilon$ is such that $\rho_{_L}(\epsilon)=1$ and $\rho_{_R}(\epsilon)=0$, or $\rho_{_L}(\epsilon)=0$ and $\rho_{_R}(\epsilon)=1$. The FCS is then given as a superposition of independent binomial laws (``partition noise''), one for each of these scattering states,
lying in an energy window $eV$, where $V$ is the voltage drop accross the barrier\cite{levitov:1993}.
 In the high temperature limit, one has to integrate the single channel result Eq.~(\ref{singleBaFCS})
over the complete Fermi-Dirac distributions in the reservoirs, given in Eq.~(\ref{eq:FD}).

From equations (\ref{Ci}), (\ref{integration}), (\ref{singleBaFCS}) and (${\cal T}=\Gamma^{(\rightarrow)}=\Gamma^{(\leftarrow)}$), one can derive the normalized noise power
$F=K_2/K_1$ in the low temperature limit and the normalized skewness $F_3=K_3/K_1$ in the high temperature limit
\cite{levitov:2001,gutman:2002cumulants}. It is interesting to note that these two quantities turn
out to be equal. 
More generally, for a mesoscopic conductor decomposed into independent channels of transparencies ${\cal T}_i$ one has~:
\begin{equation}
F(eV\gg k_{B}T)=F_3(eV\ll k_{B}T)=\frac{\sum{{\cal T}_i(1-{\cal T}_i)}}{\sum{{\cal T}_i}}
\label{F2F3}
\end{equation}
The physical information contained in the third cumulant at high temperature\cite{levitov:2001,gutman:2003HighTemp} is the same at the one contained in the
low temperature second cumulant.

Eq.~(\ref{F2F3}) will be directly checked for the mesoscopic systems considered in the rest of this paper.

\subsection{Double barriers}

For single barriers, the agreement between the exclusion model and mesoscopic models is not surprising since the boundary conditions (injection from the electrodes,...) are identical. For double barriers, it is crucial to account properly both for the boundary conditions and for Pauli exclusion principle inside the conductor.

Double barriers have been widely studied because a rich behavior results from the interplay of various effects including Pauli exclusion principle, Coulomb interactions, inelastic processes and quantum resonance\cite{blanter:2000}. In this section, we first present some results on a generic double barrier. Then we see how this system relates to various experimental devices (quantum dots, hopping on localized states, islands and wells) and how the Coulomb interaction between electrons can be introduced to account for charging effects. The case of a partly or fully diffusive island between two tunnel barriers is addressed in section~\ref{multiple}. 


\textbf{Generic double barrier}

\begin{figure}
\centerline{\includegraphics[width=7cm]{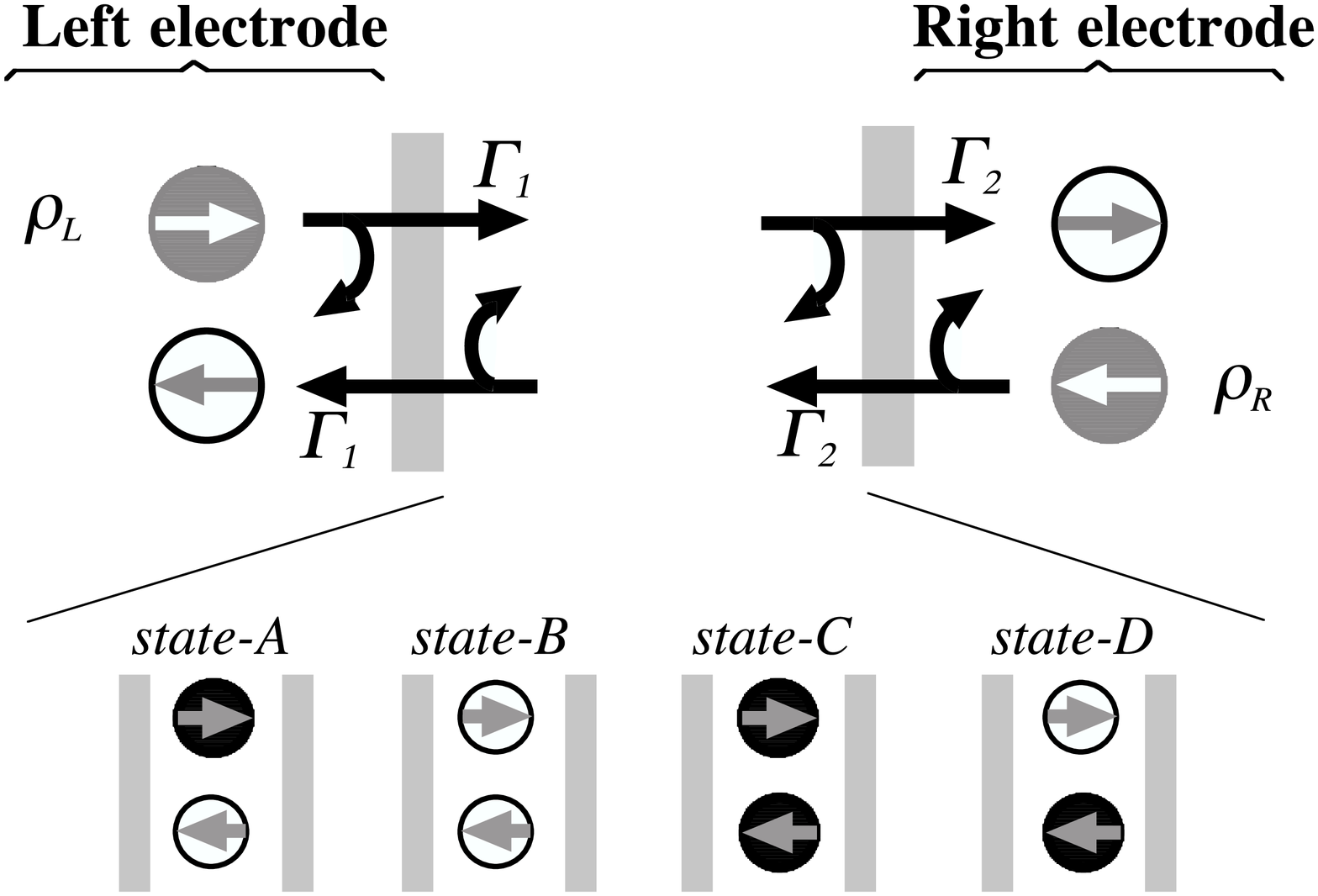}}
\caption{Upper Fig.~: Double symmetrical barriers (counter-flows model with $N=1$, $=\Gamma_{i}^{(\rightarrow)}=\Gamma_i$). For legibility, the reflection probabilities $1-\Gamma_1$ and $1-\Gamma_2$ are not written. Lower Fig.~: The internal states of the system.
}
\label{fig:doubleBarrier}
\end{figure}

We first consider two symmetric barriers of transmission $\Gamma_1$ and $\Gamma_2$, temporarily in the zero temperature limit $eV\gg k_{B}T$ ($\rho_{_L} =1$ and $\rho_{_R} =0$). The upper graph of Fig.\ref{fig:doubleBarrier} depicts the corresponding counter-flows model, with $N=1$, while the lower graph labels the $2^{2N}$ internal states of the system.

If the charge counting is done over the second barrier (arbitrary choice), the counting matrix $M_z$ is~:
\begin{equation}
M_z=
\left( \begin{array}{cccc}
\Gamma_1 \Gamma_2 z & \Gamma_1 & \Gamma_2 z & 1 \\
(1-\Gamma_1) \Gamma_2 z  & 1-\Gamma_1  & 0    & 0\\
\Gamma_1 (1-\Gamma_2) & 0 & 1-\Gamma_2 & 0\\
(1-\Gamma_1) (1-\Gamma_2) & 0   & 0     & 0
\end{array} \right)
\end{equation}

In this matrix, the states are ordered from state-A (upper-left) to state-D (lower-right).
The eigenvalues of $M_z$ can be easily found and the cumulant generating function ${\cal S}_{t}(z)$ is proportional to the logarithm of the largest one~:
\begin{equation}
{\cal S}_{t}(z)/t=\ln \left( { 1 - \frac{ \Gamma_1 + \Gamma_2 - \Gamma_1\Gamma_2 z}{2}} + 
  {\sqrt{ 
      {{{\left( 1- \frac{ \Gamma_1 + \Gamma_2 - \Gamma_1\Gamma_2 z }{2} \right) }^2}}
-\left( 1 - \Gamma_1 \right) 
         \left( 1 - \Gamma_2 \right)
} } \right)
\label{StdoubleBa}
\end{equation}
The symmetry of this expression 
between $\Gamma_1$ and $\Gamma_2$ illustrates that  the charge counting can be performed on any side of the system without changing the result. As expected, the single barrier FCS is recovered if one barrier is transparent 
($\Gamma_1 \mathrm{ or } \Gamma_2=1$). Eq.~(\ref{StdoubleBa}) extends two other results first obtained by de Jong in the Boltzmann-Langevin formalism~: the first and second cumulants of a double barrier\cite{deJong:1995} and the FCS of a double tunnel barrier ($\Gamma_1\Gamma_2 \ll 1$)~\cite{deJong:1996double}. It would be interesting to relate  these classical
expressions to those of a full quantum treatment of the double barrier system. As described below, such a precise
connection may be established in the tunneling regime where both transmissions $\Gamma_{1}$ and
$\Gamma_{2}$ are very small. 

For arbitrary fillings $\rho_{_L}$ and $\rho_{_R}$ of the electrodes, the counting matrix $M_z$ has no zero element and the eigenvalue problem is still manageable but more tedious. The power expansion method presented in section~\ref{methode} is chosen to derivate the cumulants $C_1$, $C_2$ and $C_3$, and integration over Fermi-Dirac distribution in the electrodes, according to Eq~\ref{integration} and Eq~\ref{FermiDirac}, gives the cumulants $K_1(eV/k_{B}T)$, $K_2(eV/k_{B}T)$ and $K_3(eV/k_{B}T)$. It is useful to define:
\begin{equation}
\Gamma_{12}=\Gamma_1+\Gamma_2-\Gamma_1 \Gamma_2
\end{equation}

One finds
\begin{equation}
C_1/t=(\rho_{_L}-\rho_{_R}) \frac{\Gamma_1 \Gamma_2}{\Gamma_{12}}
\end{equation}
\begin{equation}
C_2/t=
\frac{\Gamma_1 \Gamma_2 \Gamma_{12}^2 (\rho_{_L}+\rho_{_R})
-(\rho_{_L}^2+\rho_{_R}^2)\Gamma_1^2\Gamma_2^2 (2-\Gamma_{12})
-2 \rho_{_L} \rho_{_R} \Gamma_1 \Gamma_2 (\Gamma_1^2+\Gamma_2^2-\Gamma_1 \Gamma_2 (\Gamma_1+\Gamma_2))}
{\Gamma_{12}^3}
\end{equation}

For $eV\ll k_{B}T$, we find $K_2/ K_1=2k_{B}T/eV$, that is the well known Johnson-Nyquist thermal noise formula, more often written $S_I=4k_{B}TG$ where the conductance $G$ and the current noise power spectral density $S_I$ are given by Eq.~(\ref{defGSI}).

\begin{figure}
\centerline{\includegraphics[width=12cm]{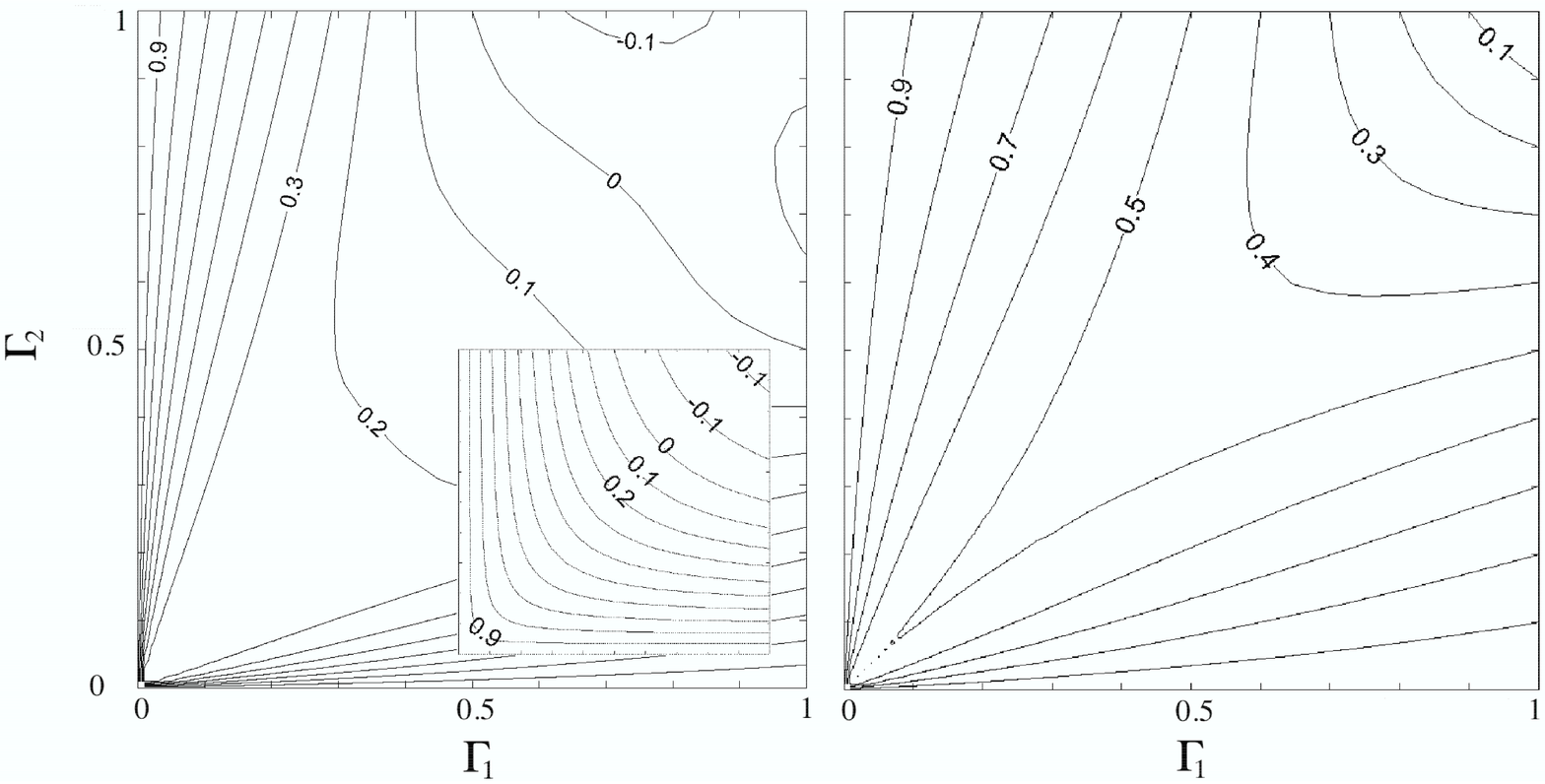}}
\caption{
Normalized skewness $F_3=K_3/K_1$ for a double symmetrical barrier. Left Fig.: Zero temperature limit ($eV\gg k_{B}T$) (insert: the exclusion principle is deactivated between the barriers). Right Fig.: High temperature limit ($k_{B}T\gg eV$). 
}
\label{fig:F3doubleBarrier}
\end{figure}

We focus now on the normalized skewness $F_3=K_3/K_1$. We give below its $eV\gg k_{B}T$ and $eV\ll k_{B}T$ limits, which -as would be expected from a quantum mechanical derivation (Eq.~(\ref{F2F3}))- is equal to the normalized noise power $F=K_2/K_1$ in the zero temperature limit. Fig.\ref{fig:F3doubleBarrier} shows that depending of $\Gamma_1$ and $\Gamma_2$, the skewness changes sign in the low temperature limit (left Fig.) and not in the high temperature one (right Fig.). The insert on the left Fig. shows $F_3=(1-\Gamma_1 \Gamma_2 / \Gamma_{12})(1-2\Gamma_1 \Gamma_2 / \Gamma_{12})$ obtained when the exclusion principle is deactivated between the two barriers. The change of sign is still observed and thus, it cannot be attributed to correlation inducted by the exclusion principle inside the conductor.

\begin{equation}
F_3(eV\gg k_{B}T)=\frac{\Gamma_1^4 + \Gamma_2^4 + \Gamma_1^3 \Gamma_2^3 (4 + \Gamma_1 + \Gamma_2) + 
  \Gamma_1^2 \Gamma_2^2 (6 - 3 \Gamma_1 - \Gamma_1^2 - 3 \Gamma_2 - \Gamma_2^2) - 
  \Gamma_1 \Gamma_2 (2 \Gamma_1^2 + \Gamma_1^3 + 2 \Gamma_2^2 + \Gamma_2^3) 
}{\Gamma_{12}^4}
\end{equation}
\begin{equation}
F_3(eV\ll k_{B}T)=1- \frac{\Gamma_1 \Gamma_2 (2-\Gamma_{12})}{\Gamma_{12}^2}=F(eV\gg k_{B}T)
\end{equation}


\textbf{Tunnel limit} 

The FCS of a double tunnel barrier can be obtained from the general expression Eq.(\ref{StdoubleBa}) for the double barrier. Nevertheless, it is interesting to derive it with the Symmetric Simple Exclusion Model with $N=1$.
For arbitrary fillings $\rho_{_L}$ and $\rho_{_R}$ of the electrodes, and charge counting done over the second barrier, the counting matrix is~:
\begin{equation}
W_z=
\left( \begin{array}{cc}
-\Gamma_2 \left(1-\rho_{_R} \right) -\Gamma_1 \left(1-\rho_{_L} \right)  & \Gamma_1 \, \rho_{_L} + \Gamma_2 \, \rho_{_R} \, z^{-1}\\
\Gamma_2 \left(1-\rho_{_R} \right)  \, z +   \Gamma_1 \left(1-\rho_{_L} \right)  & -\Gamma_1 \, \rho_{_L} - \Gamma_2 \, \rho_{_R}
\end{array} \right)
\end{equation}
Following section~\ref{methode}, the cumulant generating function ${\cal S}_{t}(z)$ is proportional to the largest eigenvalue of $W_z$. After a few lines of algebra, we find~:
\begin{center}
\begin{displaymath}
{\cal S}_{t}(z)/t=
- \frac{ \Gamma_1+\Gamma_2}{2} + 
{\sqrt{
  { \left( \frac{\Gamma_1 + \Gamma_2 }{2} \right) ^2}
+  \Gamma_1\Gamma_2 \left( \rho_{_L}\left(1-\rho_{_R}\right) \left( z-1 \right)
+ \rho_{_R}\left(1-\rho_{_L}\right) \left( z^{-1}-1 \right) \right)
}}
\end{displaymath} 
\begin{equation}
\label{above}
\end{equation}
\end{center}

We recover the FCS derived in\cite{bagrets:2002}, which slightly differs from the one derived in\cite{Belzig2002}.


\textbf{Experimental systems} 

Depending on the elastic or inelastic nature of the barriers, several double barrier systems are traditionally considered. If we restrict ourselves to elastic barriers, the size of the island between the barriers is another source of experimental diversity.  

In large but finite-size islands (``quantum island''), the quasi energy levels 
are discrete and the generic system described above can model each such level. 
At this stage, it may be useful to describe in more detail the connection between
a microscopic quantum coherent model for a single conducting channel in the
presence of a double barrier in the tunneling limit, and the corresponding SSEP.
At zero temperature, the generating function ${\cal S}_{t}(z)/t$ for the microscopic
quantum model is given by\cite{levitov:1993}~:
\[{\cal S}_{t}(z)/t=v_{F}\int_{k_{\mathrm{min}}}^{k_{\mathrm{max}}}\frac{dk}{2\pi}\ln({\cal T}(k)(z-1)+1),\]
where ${\cal T}(k)$ is the transmission coefficient for an incoming electronic 
wave-function with wave-vector $k$ and $k_{\mathrm{max}}-k_{\mathrm{min}}=\frac{eV}{\hbar v_{F}}$,
$v_{F}$ being the Fermi velocity in the electrodes.
For a fully coherent system, ${\cal T}(k)$ has to obey the quantum series composition
rule for transmission matrices. In the limit of small transmissions ${\cal T}_{1}$ and ${\cal T}_{2}$,
this yields the well-known Lorentzian resonance profile:
\[{\cal T}(k)=\frac{4{\cal T}_{1}{\cal T}_{2}}{({\cal T}_{1}+{\cal T}_{2})^{2}+16(k-k_{0})^{2}l^{2}},\]
where $l$ is the distance between the two barriers. In the limit where
$eV/\hbar$ is much larger than the decay rate $\gamma=\frac{v_{F}}{2l}({\cal T}_{1}+{\cal T}_{2})$
of the resonant level in the central region, and if the average energy of this level is such that
$k_{0}$ falls in the interval $[k_{\mathrm{min}},k_{\mathrm{max}}]$, we may take
the $k$ integration over the whole real axis, yielding exactly Eq~(\ref{above})
in the special case $\rho_{_L}(\epsilon)=1$ and $\rho_{_R}(\epsilon)=0$,
provided we set $\Gamma_{i}=\frac{v_{F}}{2l}{\cal T}_{i}$ ($i=1,2$)\cite{deJong:1996double}.
So in the tunnel limit, the simple SSEP may be viewed as the result of integrating
the contributions of quantum-mechanically coherent electrons over an energy window
larger than the width $\hbar \gamma$ of a discrete level inside the
cavity delimited by the two barriers.

When the island is infinite is the transverse direction (``Quantum well''), the energy levels become a continuous energy band~: the generic model still applies but the integration Eq.~(\ref{integration})  should be done versus wavevectors\cite{blanter:2000} $k_z$ (longitudinal) and $k_\bot$ (transverse direction) rather than versus the energy $\epsilon$.

Finally, when the island is small (``Quantum dot'') or when it consists in localized states (dopants, impurities...), the energy levels are well separated but one can no longer neglect that electrons entering or leaving the island are changing its Coulomb electrostatic energy, which induces a shift of the energy levels (``charging effect''). We illustrate in the simple example below how to account for such an effect.

\begin{figure}
\centerline{\includegraphics[width=7cm]{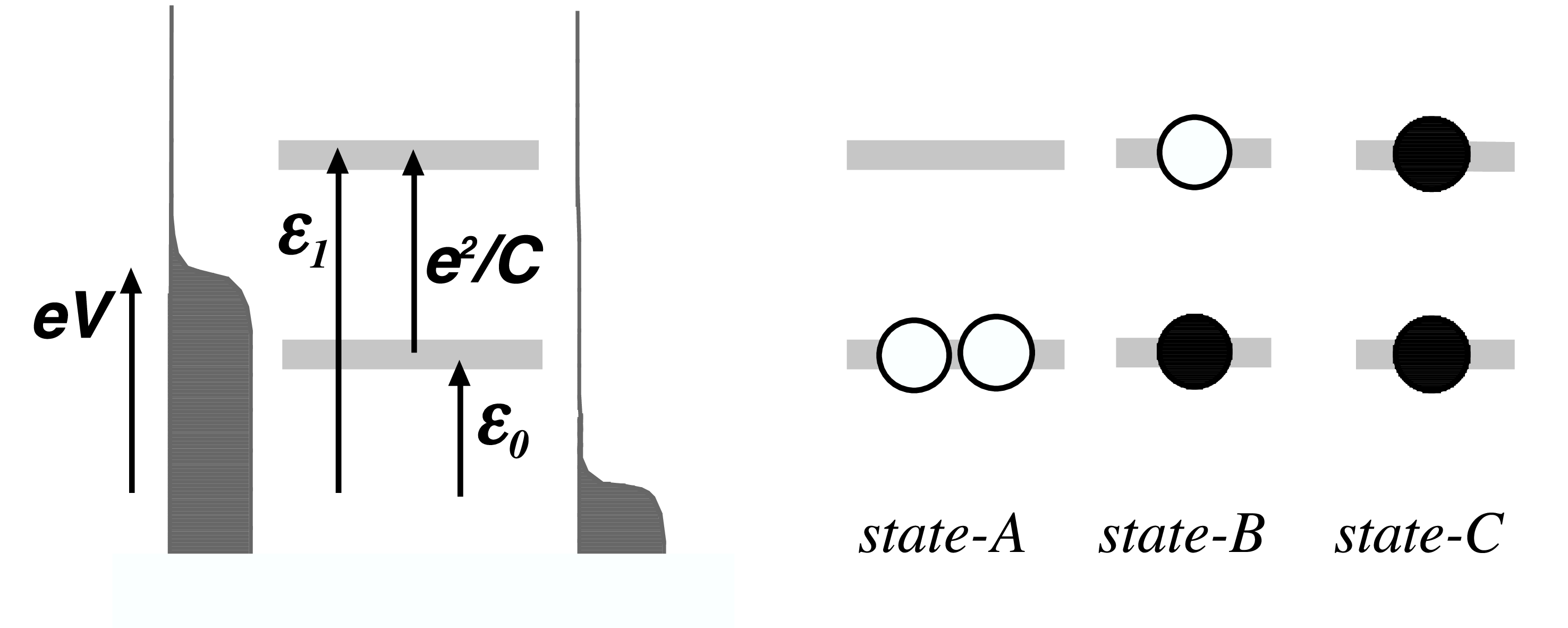}}
\caption{Left Fig.: Energy levels associated with the two sites in interaction. Right Fig.: The three internal states of the system. The black disks are electrons and the circles are available sites.}
\label{fig:niveauxcharging}
\end{figure}

\textbf{Charging effect} 

\begin{figure}
\centerline{\includegraphics[height=12cm,angle=90]{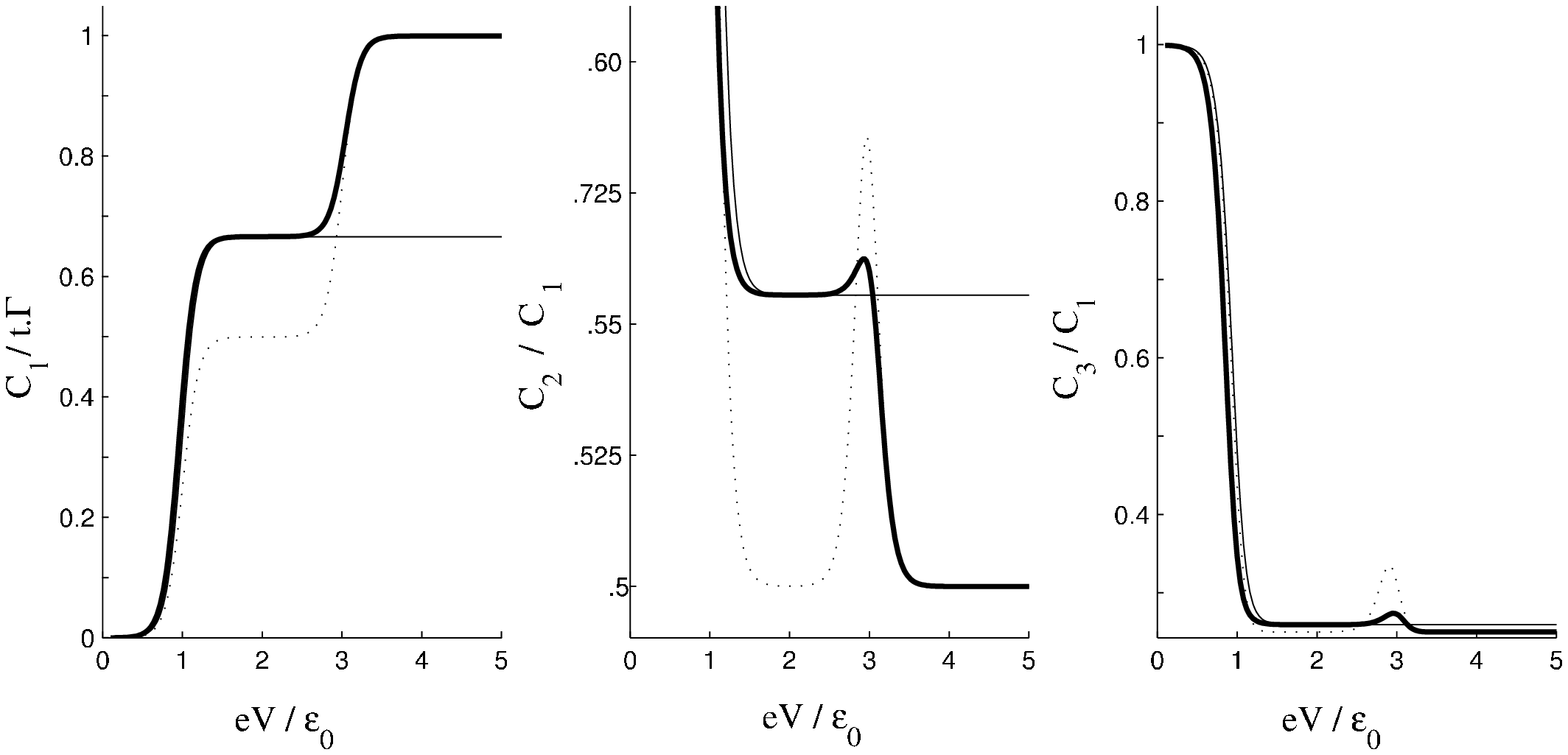}}
\caption{Two sites interacting via the Coulomb electrostatic repulsion and tunnel coupled to electrodes (see text). Left fig.: Normalized current $C_1/ t \Gamma$. Central fig.: Normalized noise power $C_2/C_1$. Right fig.: Normalized current skewness $C_3/C_1$. Thick line~: present model, thin and dash lines~: alternative models for comparison (see text).}
\label{fig:charging}
\end{figure}

We consider now a localized state tunnel-coupled to two electrodes with Fermi-Dirac fillings $\rho_{_L}(\epsilon)$ and $\rho_{_R}(\epsilon)$  (Fig. \ref{fig:niveauxcharging}). The coupling is elastic (for example, it could be due to some overlapping of the localized state wavefunction with electrodes) and characterized by two symmetrical transmission coefficients $\Gamma_1=\Gamma_2=\Gamma$ with no dependence with $\epsilon$. Up to two electrons can be trapped on the localized state (spins up and down for example). If no electron is trapped, the two available sites are degenerated at energy level $\epsilon_{0}$. The trapping of one electron shifts to $\epsilon_{1}=\epsilon_{0}+e^2/C$ the energy level of the remaining empty site, where $C$ is the effective capacitance of the island. This system has three internal states, corresponding to $0$, $1$ or $2$ trapped electrons (States A, B and C on Fig.\ref{fig:niveauxcharging}) and the corresponding counting matrix is
\begin{equation}
W_z=
\Gamma
\left( \begin{array}{ccc}-2 \rho_{_L}(\epsilon_{0})  - 2 \rho_{_R}(\epsilon_{0})       &  (1-\rho_{_L}(\epsilon_{0}))  + (1-\rho_{_R}(\epsilon_{0}))  z & 0  \\
 2 \rho_{_L}(\epsilon_{0})  + 2 \rho_{_R}(\epsilon_{0})  / z  & -2+\rho_{_L}(\epsilon_{0})-\rho_{_L}(\epsilon_{1}) +\rho_{_R}(\epsilon_{0})-\rho_{_R}(\epsilon_{1})  & 2(1-\rho_{_L}(\epsilon_{1}))+2(1-\rho_{_R}(\epsilon_{1})) z \\
0 & \rho_{_L}(\epsilon_{1})  + \rho_{_R}(\epsilon_{1})  /z & -2+\rho_{_L}(\epsilon_{1})+\rho_{_R}(\epsilon_{1}))\end{array} \right)
\end{equation}
With a software like Mathematica, the analytical equations for the cumulants $C_1$, $C_2$ and $C_3$ are easily derived by the series expansion method (section~\ref{methode}). These expresssion are not reported here but have been used to generate the normalized cumulants of Fig.\ref{fig:charging}  (Thick lines) for parameters $\epsilon_0 / k_{B}T=10$, $(e^2/C) / k_{B}T = 20$ and $\Gamma\ll1$. The first cumulant $C_1$ is normalized by the averaged transfer charge $2.t.\Gamma/2$ expected in the high driving voltage limit.
For comparison, alternative models are shown on the same figure. The dashed lines correspond to 2 sites at fixed energy levels $\epsilon_0$ and  $\epsilon_1$ and connected to the electrodes through  tunnel barriers of transmission $\Gamma$. This modeling of the original problem is expected to  account for the high $eV$ limit ($\epsilon_1\ll eV$). The thin line corresponds to 1 site between tunnel barriers of transmissions $2\Gamma$ (left barrier) and $\Gamma$ (right barrier). This model is expected to account for the intermediate plateau regime for which only one site of the localized state contributes efficiently to transport ($\epsilon_0\ll eV\ll \epsilon_1$). Fig.\ref{fig:charging} shows that the general solution to the first problem (thick line) is in good agreement with the two limiting cases considered for comparison.


\subsection{Multiple barriers~: from double wells to diffusive islands between tunnel contacts\label{multiple}}

In this section we discuss the general SSEP model presented at the end of section~\ref{tunnelExclusion} ,  $(N-1)$ identical barriers of transmission $\Gamma \ll 1$ are sandwiched between two tunnel barriers of transmission $\Gamma_{_L}$ and $\Gamma_{_R}$. 
Writing down that the mean current and its variance do not depend on the barrier over which they are measured, we showed in\cite{derrida2004} how to derive the first two cumulants~~:
\begin{equation}
C_1/t= \frac{\Gamma}{N_{1}} \left( \rho_{_L}-\rho_{_R} \right) 
\end{equation}

\begin{equation}
C_2/t= \frac{\Gamma}{N_{1}} \left[
 \left( \rho_{_L}+\rho_{_R}-2 \rho_{_L}  \rho_{_R} \right)
-  \frac{2}{3} {\left( \rho_{_L}-\rho_{_R} \right)}^2  
\left(
1
- \frac{1}{2N_{1}} 
- \frac{\lambda}{2{N_{1}}^2 \left( N_{1} - 1 \right) }
\right)
\right]
\end{equation}

with $N_{1}=N-1+\frac{\Gamma}{\Gamma_{_L}}+\frac{\Gamma}{\Gamma_{_R}}$ and $\lambda=\frac{\Gamma}{\Gamma_{_L}} \left( \frac{\Gamma}{\Gamma_{_L}}-1\right)  \left( 2\frac{\Gamma}{\Gamma_{_L}}-1\right)  
+ \frac{\Gamma}{\Gamma_{_R}} \left( \frac{\Gamma}{\Gamma_{_R}}-1\right)  \left( 2\frac{\Gamma}{\Gamma_{_R}}-1\right)$

For arbitrary $N_1$, the third cumulant $C_3$ can be derived from the third order series expansion of the cumulant generating function given in\cite{derrida2004} (Eq.~(C.14)). We give below the normalized skewness in the large $N_1$ (or $N$) limit.

\textbf{Experimental systems} 

Three limiting cases of the above general formula could be
relevant to mesophysics.

$\bullet$ Firstly, for N=2, the system is a triple barrier with three different tunnel transmissions. Such systems have been widely studied as a model for double localized state, coupled quantum dot and double well structures (for example see\cite{blanter:2000,kinkhabwala:2000}). It would be interesting to compare our predictions to the experimental results.

$\bullet$ In the limit $N\rightarrow \infty$ for constant $\Gamma$, $\Gamma_{_L}$ and $\Gamma_{_R}$, the relative contribution of the two out-most barriers is vanishingly small and we obtain a diffusive medium in good contact with the electrodes. Section~\ref{diffusive} addresses this purely diffusive regime and gives its FCS. At this point, we just point that $C_1$ and $C_2$ are in agreement with the one found by alternative condensed-matter formalisms\cite{nagaev:1992,beenakker:1992,nazarov:1994,altshuler:1994,deJong:1995,blanter:1997,sukhorukov:1998}.

$\bullet$ The limit $N\rightarrow \infty$ for constant $\Gamma$, $N \Gamma_{_L}$ and $N\Gamma_{_R}$ accounts for a diffusive island between two tunnel contact,the resistances of which  are comparable to the resistance of the diffusive island itself. We define as $q_{_L}=\Gamma / (N \Gamma_{_L})$ and $q_{_R}=\Gamma / (N \Gamma_{_R})$ the ratios of the left and right contact resistances over the diffusive island resistance. The large $q_{_L}$ and $q_{_R}$ limit accounts for simple or double tunnel barriers while vanishing values of $q_{_L}$ and $q_{_R}$ account for a purely diffusive medium.

The normalized noise power $F$ (Eq.~\ref{FanoFactorGene}), is found to be, for an arbitrary temperature~:
\begin{equation}
F=\frac{K_2}{K_1}=\frac{2k_{B}T}{eV}(1-s)+s \coth(\frac{eV}{2k_{B}T})
\label{tunnelDiff}
\end{equation}
with
\begin{equation}
s=\frac{1}{3} +\frac{2 \left(q_{_L}^3+q_{_R}^3 \right)}{3 \left(1+q_{_L}+q_{_R}\right)^3} 
\end{equation}
The well known noise power of double tunnel barriers\cite{Chen:1991,Davis:1992,blanter:2000} and diffusive media\cite{nagaev:1992,beenakker:1992,nazarov:1994,altshuler:1994,deJong:1995,blanter:1997,sukhorukov:1998} are recovered in the large and small $q_{_L}$, $q_{_R}$ limits. The normalized skewness is found to be~:


\begin{equation}
F_3=\frac{K_3}{K_1}=1 + \frac{k_{B}T}{eV} 3\left( s-1-2f \right) \coth (\frac{eV}{2k_{B}T})   + 
  {  \frac{ f-3(s-1)/2 +2f { \left[ \cosh(\frac{eV}{2k_{B}T}) \right] } ^2}{{ \left[ \sinh (\frac{eV}{2k_{B}T}) \right] }^2}  }
\label{F3_general}
\end{equation}
with 
\begin{equation}
f=-\frac{3}{10} +3 s \left( s - 1/2 \right)  -\frac{6 (q_{_L}^5+q_{_R}^5) }{5  {(1+q_{_L}+q_{_R})^5}   }
\end{equation}

Eq.~(\ref{F3_general}) bridges continuously between two simple situations : double tunnel barriers (large $q_{_L}$, $q_{_R}$) and diffusive media (small $q_{_L}$, $q_{_R}$) for which we also recover known results\cite{gutman:2002cumulants,nagaev:2002}. In the low and high temperature limits, we found~:
\begin{equation}
   F_3(eV\gg k_{B}T)=1+2f    \,\,\,\,\,\,\,\,\,\,\,\,\,\,\,\,\,\,\,\, F_3(eV\ll k_{B}T)=s=F(eV\gg k_{B}T)
\end{equation}
The high temperature skewness is equal to the low temperature noise power, as would be expected from a
fully quantum treatment (see Eq.~(\ref{F2F3})).
\subsection{Diffusive medium\label{diffusive}}

In this section, we consider a diffusive medium with good contacts to the electrodes. 
We first consider the SSEP model in the $N\rightarrow \infty$ limit, which is enough to account for a purely diffusive behavior. At the end of this section, we consider the cross-over from a ballistic to diffusive conduction. To do so, the counter-flows model will be required.

\begin{table}
\begin{center}
\begin{tabular}{|c|c|c|c|}
\hline
 &   &  $K_n/K_1$& $K_n/K_1$\\
n &  $Normalized~~cumulants~:~~~\frac{C_n}{ (t\Gamma/N)}$ &for&$for$\\
 &   & {\bf $k_{B}T\ll eV$} & {\bf $k_{B}T\gg eV$}\\
\hline
1 & $X$ & $1$ & $1$\\ 
2 & ${\displaystyle \left[ 3 - {X^2} - 3{Y^2} \right] / 6}$ & ${\frac{1}{3}}$ & $\frac{2k_{B}T}{eV}$\\ 
3 & ${\displaystyle {X^3}/15} + X{Y^2}$ & ${\frac{1}{15}}$ & $\frac{1}{3}$\\ 
4 & ${\displaystyle  \left[ -9{X^4} + {X^2}\left( 7 - 462{Y^2} \right)  -  105{Y^2}\left( -1 + {Y^2} \right) \right] / 210}$
 & $-{\frac{1}{105}}$ & ${\frac{2k_{B}T}{3eV}}$\\ 
5 & ${\displaystyle 4{X^5}/105} +   X{Y^2}\left( -3 + 4{Y^2} \right)  +   {\displaystyle {X^3}\left( -1 + 120{Y^2} \right) /21}$ & $-{ \frac{1}{105}}$ & $-\frac{1}{5}$\\ 
6 &   $  [ -20{X^6} - 33{X^4}  \left( -1 + 244{Y^2} \right)  -  231{Y^2}\left( 3 - 7{Y^2} + 4{Y^4} \right)  -$  &  &   \\
 & $      11{X^2}\left( 1 - 618{Y^2} + 1044{Y^4} \right)  ] / 462 $          & ${\frac{1}{231}}$ & $-{\frac{2k_{B}T}{5eV}}$   \\
 \hline
\end{tabular}
\caption{First cumulants $C_n$ for diffusive medium for arbitrary filling of the electrodes, and low and high temperature normalized cumulants $K_n/K_1$.  ($X=(\rho_{_L}-\rho_{_R})$ and $Y=(\rho_{_L}+\rho_{_R}-1)$).}
\end{center}
\end{table}

\textbf{SSEP diffusive medium} 

In our previous paper\cite{derrida2004}, we detail the derivation of the FCS for SSEP models in the $N\rightarrow \infty$ limit and we shall only summarize the derivation here. The cumulant generating function ${\cal S}_t$ depends on $N$, $z$, $\rho_L$ and $\rho_R$ but to first order in $1/N$,   ${\cal S}_t$ turns out to depend only on a combinaison $\omega$ of these parameters~:
\begin{equation}
\omega=  {(z-1) (\rho_L z -\rho_R - \rho_L \rho_R(z-1) )\over z}
\end{equation}
Consequently, to first order in $1/N$ one can write~:
\begin{equation}
{\cal S}_t /t= {\Gamma \over N} R(\omega)
\end{equation}
where the expression of $R(\omega)$ has been conjectured. Thus, if the FCS is known for two arbitrary fillings of the electrodes, then it can be deduced for any fillings or at any temperature.

Based on the exact derivation of the first 4 cumulants, we conjectured that for $\rho_{_L}=\rho_{_R}=1/2$, the statistics of current fluctuations is Gaussian (at order 1/N). This fully determines $R(\omega)$ and thus the current statistics at a arbitrary fillings  $\rho_{_L}$ and  $\rho_{_R}$~:
\begin{equation}
{\cal S}_t /t= {\Gamma \over N} \left[  \log(\sqrt{1+\omega}+\sqrt{\omega}) \right]^2
\end{equation}

This expression is identical to the one found at zero temperature by\cite{lee} and at arbitrary temperature by\cite{gutman:2002cumulants} (and by\cite{nazarov:1999} with minor discrepancies). Tables 2 gives the first cumulants $C_n$, $K_n/K_1$ for $k_{B}T\ll eV$ and for $k_{B}T\gg eV$. Right/Left symmetry and particle/hole symmetry suggest the change of parameters~Ê: $X=(\rho_{_L}-\rho_{_R})$ and $Y=(\rho_{_L}+\rho_{_R}-1)$. Agreement is found with the first four cumulants which can be easily derived from\cite{nagaev:2002}. 

\begin{figure}
\centerline{\includegraphics[angle=90, width=9cm]{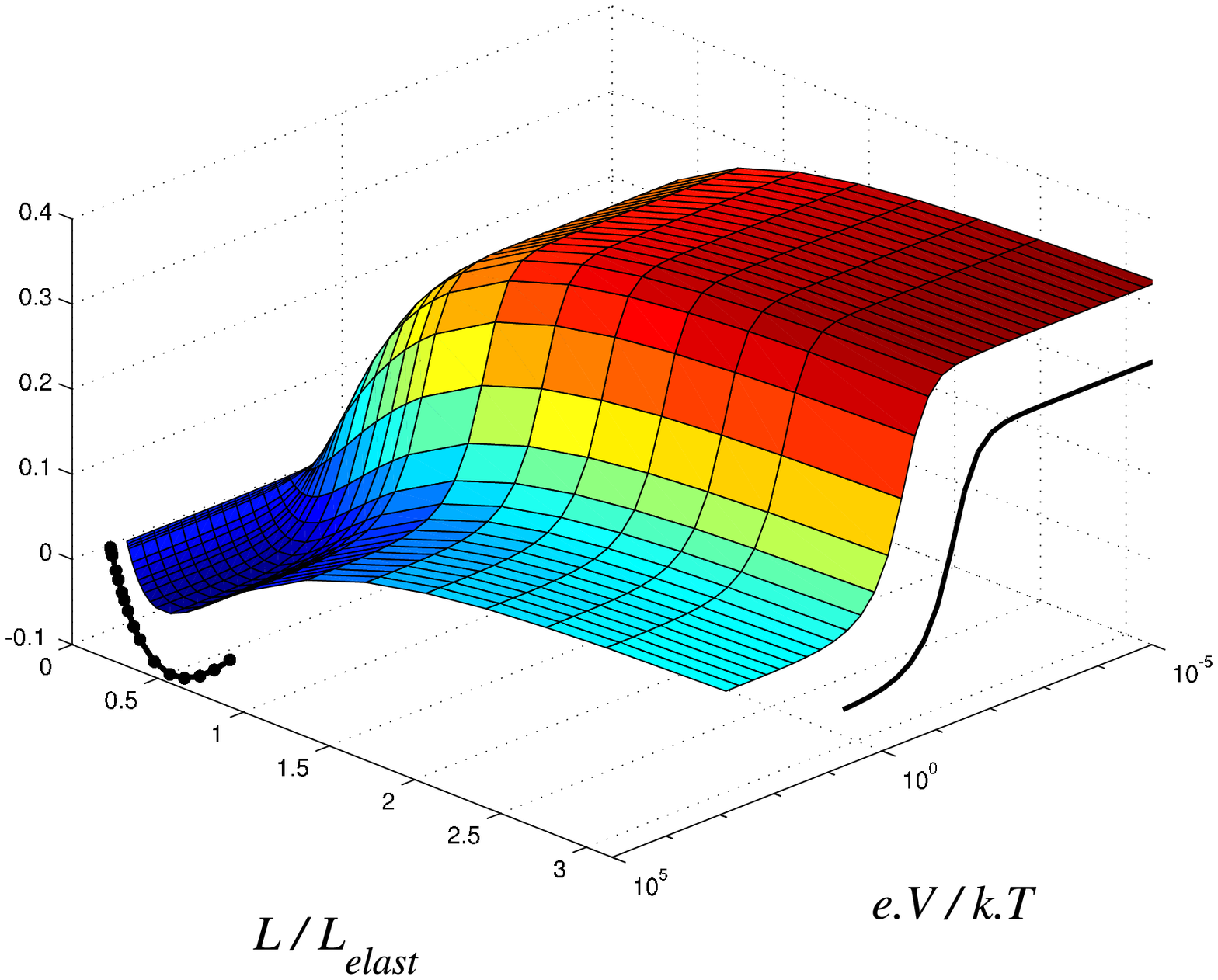}}
\caption{Normalized current skewness $K_3/K_1$ in a ballistic/diffusive medium parametrized by the ratio $L/L_{elast}$ of the conductor's length $L$ over a elastic mean free path $L_{elast}$ (counter-flows with $N=8$). The two lines correspond to known limit for $L/L_{elast}\gg 1$ (continuous line) and for $L/L_{elast}\ll 1$ with $eV\gg k_{B}T$ (pearly line). }
\label{fig:crossOver}
\end{figure}

\textbf{Ballistic-Diffusive cross-over} 

We now consider a counter-flows model with uniform symmetrical transmission coefficients $\Gamma$ in the bulk of the conductor ($\Gamma=\Gamma_{i}^{(\rightarrow)}=\Gamma_{i}^{(\leftarrow)}$ for $1<i<N+1$ ), unity transmission at the boundaries ($\Gamma_{1}^{(\leftarrow)}=\Gamma_{1}^{(\rightarrow)}=\Gamma_{N+1}^{(\leftarrow)}=\Gamma_{N+1}^{(\rightarrow)}=1$), and we consider the $N\rightarrow \infty$ limit taken at constant $(1-\Gamma)(N-1)$. We can define the mean free path $L_{elast}$ of a single electron as $1/(1-\Gamma)$ and the length $L$ of the conductor as $N-1$. Then  $(1-\Gamma)(N-1)=L/L_{elast}\ll 1$ corresponds to a ballistic conductor while $L/L_{elast} \gg 1$ to a diffusive one\cite{deJong:1996book,liu}. We explore the normalized current skewness $F_3=K_3/K_1$ versus both the ballistic-diffusive cross-over and the thermal ($k_{B}T\gg eV$) - shot noise ($eV\gg k_{B}T$) cross-over. At zero temperature, it was shown\cite{roche:2002exclusion} that $N$ as small  as $8$ gives a correct picture of the skewness, even in the diffusive limit. We performed a numerical simulation of such a small $N$ by solving numerically the largest eigenvalue of the counting matrix. Fig.\ref{fig:crossOver} shows $F_3(L/L_{elast},eV/k_{B}T)$. For comparison, the continuous line is calculated from the analytical results in the diffusive limit and good agreement is found with the numerical estimation. 
This is an interesting result, which shows a tendency of the simple exclusion model to reproduce
the same FCS as for a larger class of models. This raises the question of trying to identify
more precisely the key features which are required for a stochastic model to yield this FCS.
The pearly line corresponds to a single scatterer with the same conductance as the system in the low temperature limit. As $L/L_{elast}\rightarrow 0$ ($L$ may be arbitrary large), 
the system is expected to converge to such a point scatterer limit and a good agreement is found with the simulation. Good agreement is also found with Monte-Carlo simulations performed along  the ballistic-diffusive cross-over at zero temperature\cite{roche:2002exclusion}.
\section{Connection to microscopic models\label{quantum}}

As we have seen in section~\ref{Application} above,
there are many instances where  a simple approach in terms of purely
classical exclusion models yields an impressive
agreement with fully quantum-mechanical computations, 
specially for diffusive systems. This raises naturally the question
of trying to derive these classical models directly from a fully
quantum mechanical description of a coherent conductor. In this section,
we shall establish such a connection for the counter-flows exclusion model,
but only in the rather special situation of a {\em single} barrier.

The basic principle of this model is to discretize the single-particle 
states of conduction electrons.
For each value of the kinetic energy, we consider two types of localized 
wave-packets, moving in either
direction (from right to left, or left to right). Each possible 
wave-packet is supposed to be 
spacially located in one of $N$ cells of width $\Delta x$ into which our 
mesoscopic conductor is subdivided.
Let us first assume a perfect conductor, with no impurities. It is then 
natural to approximate
the true quantum mechanical evolution operator describing a single 
electron during a time
interval $\Delta t=\Delta x/v_{F}$ ($v_{F}$ is the Fermi velocity of 
electrons, taken to be also
the group velocity of these wave-packets) by the following matrix $U$ 
acting as:
\begin{eqnarray}
U|i,+\rangle &=& |i+1,+\rangle \\
U|i,-\rangle &=& |i-1,+\rangle 
\end{eqnarray}
Here, $|i,\pm\rangle$ denotes a wave-packet with velocity $\pm v_{F}$ 
located in the cell labelled by the 
integer $i$. From now on, we shall assume $i$ ranges over all integers, 
ignoring what happens at
the boundaries of the sample. In fact, rather than taking the 
reservoirs
explicitely into account here, we have in mind an evolution of the system 
in a finite
time interval, from an initial state with a finite number of particles, 
located on the left and on the right of the finite length region where
impurity scattering may occur. After all, reservoirs always contain a 
finite number
of particles in real experiments.
It is straightforward to check that $U$ thus defined is unitary.
States with $p$ electrons located in cells $i_{1}<i_{2}<...<i_{p}$ and 
moving
with velocities $v_{1},v_{2},...,v_{p}$ ($v_{i}=\pm$)
are obtained as usual by forming fully antisymmetrized tensor products of 
the 
single particle states $|i_{1},v_{1}\rangle, 
|i_{2},v_{2}\rangle,..., |i_{p},v_{p}\rangle$,
denoted by 
$|i_{1},v_{1};i_{2},v_{2};...;i_{p},v_{p}\rangle$.
The unitary operator $U$ allows to construct the corresponding evolution
operator $U^{(p)}$ in the space of $p$ electron states, and this latter 
operator is defined by:
\begin{equation}
\label{evolibrep}
U^{(p)}|i_{1},v_{1};i_{2},v_{2};...;i_{p},v_{p}\rangle=
|i_{1}+v_{1},v_{1};i_{2}+v_{2},v_{2};...;i_{p}+v_{p},v_{p}\rangle
\end{equation}
Now, if we prepare such a system at time $t$ in a density matrix $\rho(t)$ 
which is diagonal 
in the above-given basis of $p$ electron states, we have $\rho(t+\Delta 
t)=U^{(p)}\rho(t)U^{(p)+}$,
and equation(~\ref{evolibrep}) shows that $\rho$ remains diagonal. This 
defines a very 
simple process, where the probability 
$\mathcal{P}(i_{1},v_{1};i_{2},v_{2};...;i_{p},v_{p};t)$
evolves according to:
\begin{equation}
\label{eq:evolP}
\mathcal{P}(i_{1},v_{1};i_{2},v_{2};...;i_{p},v_{p};t+1)=
\mathcal{P}(i_{1}-v_{1},v_{1};i_{2}-v_{2},v_{2};...;i_{p}-v_{p},v_{p};t)
\end{equation}
Note that time is now written in units of $\Delta t$, which simplifies the 
notation.

The next step is to include a single barrier, located between cells $i$ 
and $i+1$.
The definition of the single particle evolution operator $U$ is now 
modified according to:
\begin{eqnarray}
U|j,+\rangle &=& |j+1,+\rangle \;\;\;(j\neq i-1)\\
U|j,-\rangle &=& |j-1,+\rangle \;\;\;(j\neq i) \\ 
U|i-1,+\rangle &=& 
\cos(\theta_{i})|i,+\rangle+\sin(\theta_{i})|i-1,-\rangle \\
U|i,-\rangle &=& \cos(\theta_{i})|i-1,-\rangle-\sin(\theta_{i})|i,+\rangle
\end{eqnarray}
There is now a finite probability $\sin^{2}(\theta_{i})$ for
the particle coming from the left (on cell $i-1$) or from the right (on 
cell $i$)
of the barrier to be backscattered. Let us denote by $\Gamma_{i}$ the 
transmission
probability, equal to $\cos^{2}(\theta_{i})$. Note that the $U$ matrix 
just written
is not the most general unitary matrix, and both transmission and 
reflection 
amplitudes are in general complex numbers. But again, we made this choice 
to keep 
a lighter notation. As before, this model for the single particle quantum 
evolution
generates also the evolution of $p$ particle states. Since the only single 
particle
states which are affected by the presence of this barrier are 
$|i-1,+\rangle$
and  $|i,-\rangle$, it is just sufficient to consider initial states where
these are both occupied. In fact, because of the Pauli principle, we have:
\begin{equation}
\label{Pauli}
U^{(2)}|i-1,+;i,-\rangle=|i,+;i-1,-\rangle
\end{equation}
We shall not write explicitely the corresponding unitary matrix $U^{(p)}$
here. But the important point to stress is that it induces an evolution
of the density matrix $\rho$ which still preserves the stability of the
subspace of diagonal matrices. The basic idea behind this property
is a description of the time evolution of the system with $p$ particles in
terms of trajectories, in the spirit of the path integral approach to
quantum mechanics. In the presence of a unique barrier, there is at most
a single path which connects an initial state 
$|i_{1},v_{1};i_{2},v_{2};...;i_{p},v_{p}\rangle$
to a final state  
$|i_{1}',v_{1}';i_{2}',v_{2}';...;i_{p}',v_{p}'\rangle$.
This holds because under successive applications of $U$, the distance of
a particle to the barrier in the initial state $|i,v\rangle$ which 
evolves
into a known final state $|i',v'\rangle$, is
uniquely determined by this final state and the total duration of this 
evolution.
So if the distances of the final locations of the particles
$i_{1}',i_{2}',...,i_{p}'$ to the barrier are all different, there is only 
one permutation
of the particles which takes the $i$'s into the $i'$'s. When two $i'$'s 
correspond
to the same distance to the barrier, the Pauli principle imposes these two 
sites
to be located on opposite sides of the barrier, and the same statement is 
true for
the $i$'s of the corresponding initial states. If we had {\em 
distinguishable}
particles, this would yield two possible paths connecting this pair of 
initial
$i$'s to this pair of final $i'$'s. But the antisymmetrization operation 
forces
us to see these two paths as a single one, as in equation~(\ref{Pauli}) 
above.

As for the pure system, we may then interpret the 
evolution of the diagonal part of the density matrix in terms of a simple 
stochastic process
for the probability 
$\mathcal{P}(i_{1},v_{1};i_{2},v_{2};...;i_{p},v_{p};t)$,
which is nothing but the counter-flows exclusion model with a single 
barrier,
defined in section~\ref{exclumod} above.

Unfortunately, it is difficult to generalize this construction to 
more than a single barrier. 
For three barriers and a single particle already, there are
in general several differents paths (generated by 
a given number of successive applications of the matrix $U$)
which connect an initial state $|i,v\rangle$
to a final state $|i',v'\rangle$. Therefore, the corresponding evolution
of the density matrix no longer preserves the subspace of diagonal 
matrices.

We took here the view that we may still obtain some valuable 
information
on the FCS by ignoring the non-diagonal elements of the density matrix.
This approximation is rather severe for a long coherent single
channel conductor with a large number of barriers, since it does not allow
to capture the phenomenon of Anderson localization. But the good agreement
obtained between purely classical models and fully quantum ones in the 
case
of a two barriers and of a diffusive medium, certainly calls for a deeper
understanding of the possible connections between these two classes of  
models.

\section{Conclusion}

In this paper, we have checked explicitely that the FCS of purely
classical stochastic models with a local exclusion constraint
coincides with the FCS of a large class of quantum mechanical 
microscopic models where the current flows between two reservoirs.
In particular, no restriction on the system size seems to be
necessary, since this coincidence appears for a small number 
of tunnel barriers, as well as in the limit of a diffusive
medium composed of an arbitrary large number of such barriers.
The key common ingredient to both types of models is the constraint
removing doubly occupied sites, imposed by the Pauli principle.

It would be very interesting to extend the classical approach
in terms of exclusion models to multi-terminal geometries.
Such generalizations have been considered in a fully quantum-mechanical
treatment of diffusive systems~\cite{blanter:1997}, or in the
semi-classical Boltzmann-Langevin approach~\cite{sukhorukov:1998}.
These works have so far focused on the correlation matrix of the
integrated charges flowing through each contact during a finite time
interval. A natural question is whether simple classical exclusion
models are able to generate these results for the second order 
cumulants. The computation of higher-order cumulants in multi-terminal
geometries is another interesting open question.

\end{document}